\documentclass[nofootinbib,pre,twocolumn,showpacs,showkeys,groupaddress,preprintnumbers,floatfix]{revtex4-1}

\usepackage{dynlearn}
\usepackage{physics}

\renewcommand{\H}{\operatorname{H}}
\renewcommand{\I}{\operatorname{I}}
\newcommand{\kB} { k_\text{B} }

\def\tbf #1 {\textbf{#1} }

\newcommand{\tikzcircle}[2][black,fill=none]{\tikz[baseline=-#2]\draw[#1,radius=#2pt] (0,0) circle ;}

\newcommand{\TCircle}{\tikzcircle{3.5}}

\newcommand{\BCircle}{\tikzcircle[blue,fill=blue]{3.5}}

\newcommand{\SubTCircle}{{\bigcirc}}
\newcommand{\TSquare}{\square}
\newcommand{\RSquare}{ {\color{red}{\square}} }

\newcommand{\StZero}{ {\color{red}{0}} }
\newcommand{\StOne}{ {\color{blue}{1}} }

\newcommand{\PPos}{x}
\newcommand{\PTyp}{s}
\newcommand{\PMem}{y}
\newcommand{\PosRV}{X}
\newcommand{\TypRV}{S}
\newcommand{\MemRV}{Y}

\definecolor{tikzgreen}{rgb}{.15,0.6,0.2}

\begin{document}

\def\ourTitle{
Variations on a Demonic Theme:\\
Szilard's Other Engines
}

\def\ourAbstract{
Szilard's now-famous single-molecule engine was only the first of three
constructions he introduced in 1929 to resolve several paradoxes arising from
Maxwell's demon. We analyze Szilard's remaining two demon models. We show that
the second one, though a markedly different implementation employing a
population of distinct molecular species and semi-permeable membranes, is
informationally and thermodynamically equivalent to an ideal gas of the
single-molecule engines. Since it is a gas of noninteracting particles
one concludes, following Boyd and Crutchfield, that (i) it reduces to a chaotic
dynamical system---called the Szilard Map, a composite of three piecewise
linear maps that implement the thermodynamic transformations of measurement,
control, and erasure; (ii) its transitory functioning as an engine that
converts disorganized heat energy to work is governed by the Kolmogorov-Sinai
entropy rate; (iii) the demon's minimum necessary ``intelligence'' for optimal
functioning is given by the engine's statistical complexity, and (iv) its
functioning saturates thermodynamic bounds and so it is a minimal, optimal
implementation. We show that Szilard's third model is rather different and
addresses the fundamental issue, raised by the first two, of measurement in and
by thermodynamic systems and entropy generation. Taken together, Szilard's suite of
constructions lays out a range of possible realizations of Maxwellian demons
that anticipated by almost two decades Shannon's and Wiener's concept of
information as surprise and cybernetics' notion of functional information.
This, in turn, gives new insight into engineering implementations of novel
nanoscale information engines that leverage microscopic fluctuations and into
the diversity of thermodynamic mechanisms and intrinsic computation harnessed
in physical, molecular, biochemical, and biological systems.
}

\def\ourKeywords{
stochastic process, hidden Markov model,
\texorpdfstring{\eM}{epsilon-machine}, causal states, mutual information,
information processing Second Law of Thermodynamics
}

\hypersetup{
  pdfauthor={James P. Crutchfield},
  pdftitle={\ourTitle},
  pdfsubject={\ourAbstract},
  pdfkeywords={\ourKeywords},
  pdfproducer={},
  pdfcreator={}
}

\title{\ourTitle}

\author{Kyle J. Ray}
\email{kjray@ucdavis.edu}

\author{James P. Crutchfield}
\email{chaos@ucdavis.edu}

\affiliation{Complexity Sciences Center and Physics Department,
University of California at Davis, One Shields Avenue, Davis, CA 95616}

\date{\today}
\bibliographystyle{unsrt}

\begin{abstract}
\ourAbstract
\end{abstract}



\preprint{\arxiv{2003.XXXXX}}

\date{\today}
\maketitle

\section{Introduction}
 
Since James Clerk Maxwell first proposed an intelligence that can violate the
Second Law via accurate observations of individual molecules and precise
control of boundary conditions, the idea has been revisited and challenged
countless times. In his 1872 book on heat, Maxwell first formally introduced
the seeming paradox: a ``finite being'' that could, in essence, capture
individual thermal fluctuations to extract macroscopic amounts of work from a
heat bath \cite{Maxw88a} in violation of the Second Law. Several years later,
William Thomson (Lord Kelvin) dubbed these beings ``Maxwell's Intelligent
Demons'' \cite{Thomson1874}. And so, the paradox of Maxwell's Demon was
born: accurate observations and precise control can overcome the Second Law of
Thermodynamics, rendering disordered heat energy into useful, ordered work.
Over the following decades, many attempted resolutions \cite{Leff02a} addressed
purely mechanical limitations imposed by how a given demon acted on its
observations to sort molecules.

Thomson makes this point quite explicitly in a lecture given before the Royal
Institution in 1879, where he closes his abstract:
\begin{quote}
The conception of the `sorting demon' is merely mechanical, and is of great
value in purely physical science. It was not invented to help us to deal with
questions regarding the influence of life and of mind on the motions of matter,
questions essentially beyond the range of mere dynamics. \cite{Thomson1879}
\end{quote}
Thomson highlights two key distinctions made in early conceptions of the demon.
First, the demon's primary task is to physically sort microscopic particles by
their individual characteristics. Second, Maxwell's Demon (MD) cannot shed
light on the influence of ``mind'' on the motion of matter. (This presumably
addressed Maxwell's and others' repeated appeals to undefined
notions such as ``intelligent beings''.)

Not until 1929, when Leo Szilard published his seminal work ``On the decrease
of entropy in a thermodynamic system by the intervention of intelligent
beings'' \cite{Szil29a}, was a direct connection established between a
thermodynamic cost and what Maxwell called intelligence---and what we now call
``information''.\footnote{Notably, Szilard discussed the manuscript's
development with Albert Einstein \cite{Lano13a}.} In this, Szilard showed that
both of Thomson's assertions could be relaxed. Notably, Szilard's constructions
do not involve the direct manipulation of individual molecules, but always
involve their observation (measurement) and control. The genius in this was to
introduce a new, operational, and minimal definition of ``mind'' as storing
information in physical states; thus, inextricably linking a demon with its
physical instantiation. While Szilard acknowledged that the biological
phenomena governing the working of a ``finite being'' were beyond the scope of
physics, he delineated the minimal capabilities a mind needed to exhibit
MD-like behavior and then created idealized machines with these abilities.
Szilard's conclusion: if the Second Law is to hold, a physical memory's
interaction with a thermodynamic system must entail entropy production.

Szilard is cited in much of the work on the subject since. A notable exception
is Landauer's 1961 article \cite{Landauer1961} that, nonetheless, responds
directly to Szilard's claim that measurement has an inherent entropy cost.
Landauer argues that the thermodynamic cost arises instead from the demon's act
to erase the measured information, necessary to ``reset'' itself to begin a new
cycle afresh. From that point forward, and for the better part of a half
century, that irreversible erasure of stored measurement information was the
source of the compensating cost was taken as the resolution of Maxwell's
paradox \cite{Benn82,Land61a}. Recent developments in information
theory and nonequilibrium thermodynamics, though, allow for more precise
accounting. The result of which is that, in the general case, measurement and
erasure act as a conjugate pair---a decrease in the cost of one increases the
cost of the other \cite{Shiz95a,Fahn96a,Bark06a,Saga12a,Boyd14b}.

What this colorful history glosses over is that Szilard's 1929 work laid out
three different constructions of thermodynamic machines. Taken together, they
were his attempt to account more generally for how the flow of heat, work, and
information (our modern word, not his\footnote{``Information'' appears only
once and, then, in a narrative sense.}) drive each step of a thermodynamic
process. In today's parlance we refer to these devices as \emph{information
engines} \cite{Boyd14b}. Since then, as history would have it, the descriptor
``Szilard Engine'' came to refer only to his first construction---the
single-molecule engine. In light of recent experimental and theoretical
developments allowing new treatments of information engines, it is pertinent to
revisit Szilard's foundational work \emph{en toto}. What additional insights
can be gleaned from the other Szilard devices, if any? How do they compare to
his first engine?

Below, we retrace Szilard's steps in constructing his second device and
investigate his reasoning using more contemporary ideas and techniques for
analyzing deterministic chaotic systems, information flow, and the energetics
of nonequilibrium thermodynamic transformations. Once completed, we turn to his
third construction that, as it turns out, is a novel view of the process of
measurement itself, when two thermodynamic systems come into contact.

\section{Demon Gas: Szilard's Second Engine}
\label{sec:SecondEngine}

Consider an ensemble of \emph{demon-particle} molecules contained in a
cylindrical tube in contact with a thermal reservoir at temperature $T$. Each
demon-particle $i = 1, \ldots, N$ is defined by two variables: a particle-type
variable $\PTyp_i \in \{\TSquare,\TCircle\}$ and a variable that relates to the
demon's knowledge $\PMem_i \in \{\StZero,\StOne\}$ about the particle type.
Demon $i$ ``knows'' its molecule's type when $\PMem_i$'s value exactly
correlates that of $\PTyp_i$. We refer to $\PMem_i$ as demon $i$'s
\emph{memory}.

Particles spontaneously convert ``monomolecularly''---Szilard's
phrasing---from one type to the other at a given rate. This rate is chosen
to maintain a particular desired equilibrium distribution $\rho_0(\PTyp)$ in
which the probability of being one type is given by $\Pr(\PTyp_i = \TSquare) =
\delta$ and the other $\Pr(\PTyp_i = \TCircle)=1-\delta $. Total particle
number $N$ is conserved. This equilibrium distribution of types can be enforced
by there being an energy difference between the particle types or, perhaps, by
spin statistics---as in the case of ortho- and para-hydrogen \cite{Path96a}.
Thus, it is not necessary that the particle-type energies differ significantly.
We assume that they do differ for the sake of generality, but not altogether
for the sake of clarity. As such, we define the $N$-particle Hamiltonian:
\begin{align}
H_0 = \epsilon_\SubTCircle N_\SubTCircle + \epsilon_\TSquare N_\TSquare
  + \sum_{i=1}^N \frac{p_i^2}{2m}
  ~,
\label{eq:Hamiltonian}
\end{align}
where $\epsilon_\SubTCircle$ and $\epsilon_\TSquare$ are the particle-type
energies ($\Delta \epsilon = \epsilon_\SubTCircle - \epsilon_\TSquare > 0$), the
particle numbers $N_\SubTCircle$ and $N_\TSquare$ sum to the total $N$, $m$ is
the particle mass, and $p_i$ the $i^{th}$ particle's momentum.

\begin{figure}
\centering
\includegraphics[width=\columnwidth]{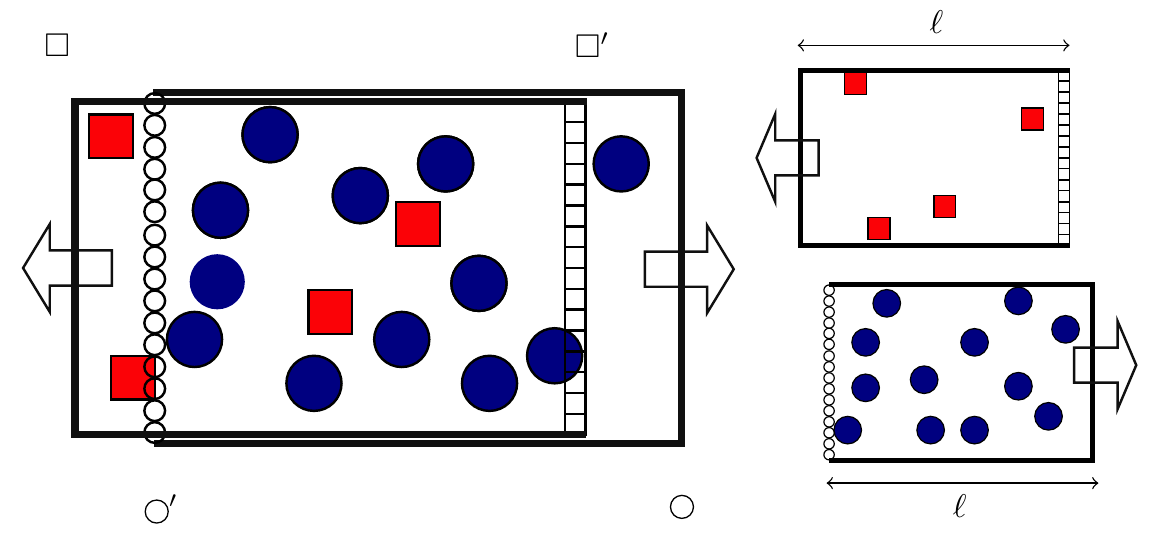}
\caption{Particle-type separation: (Left) Two constant-volume cylinders of
	length $\ell$ slide through each other, blue-circle particles $\BCircle$
	are moved from the original left ($L$) volume to the right and red-square
	particles $\RSquare$ are unaffected. A membrane permeable to $\TCircle$s
	($\TSquare$s) is depicted as a line of squares (circles), as they are, in
	essence, walls for the $\TSquare$s ($\TCircle$s) only. (Right) Individual
	cylinders at end of the particle-separation process.
	}
\label{fig:Control_SeparationSliding} 
\end{figure}

The cylinder walls are impermeable to either type of particle, and there are four pistons inside. Two of these are also impermeable---paralleling Szilard, we also denote them $\TCircle$ and $\TSquare$---and are initially set a distance $\ell$ apart. (The reason for reusing type labels as piston labels will become clear.) The other two pistons, $\TCircle'$ or $\TSquare'$, are permeable to only one of the two particle types. Each is set just inside of the impermeable pistons: $\TCircle'$ being set next to $\TSquare$ and $\TSquare'$ next to $\TCircle$. Refer to Fig. \ref{fig:Control_SeparationSliding}. These pistons are placed so they can slide one within the other, keeping the distance between $\TCircle$ and $\TCircle'$ and $\TSquare$ and $\TSquare'$ fixed at $\ell$. We can think of this system as two overlapping cylinders of fixed length that can slide relative to each other; each having an impermeable wall ($\TCircle$ or $\TSquare$) at one end and a semi-permeable membrane ($\TCircle'$ or $\TSquare'$) at the other.

Szilard specifies a cyclic control protocol with three key transformations:
\emph{Measurement}: in which each particle's initial type is stored in its
memory; \emph{Control}: in which the system's thermodynamic resources are
manipulated; and \emph{Erasure}: in which the measurements are leveraged to
return the overall system to its initial configuration. These steps generally
outline the behavior of information engines as they leverage information
resources to gain thermodynamic advantage; cf. Ref. \cite{Boyd14b}.

\begin{figure}
\includegraphics[width=\columnwidth]{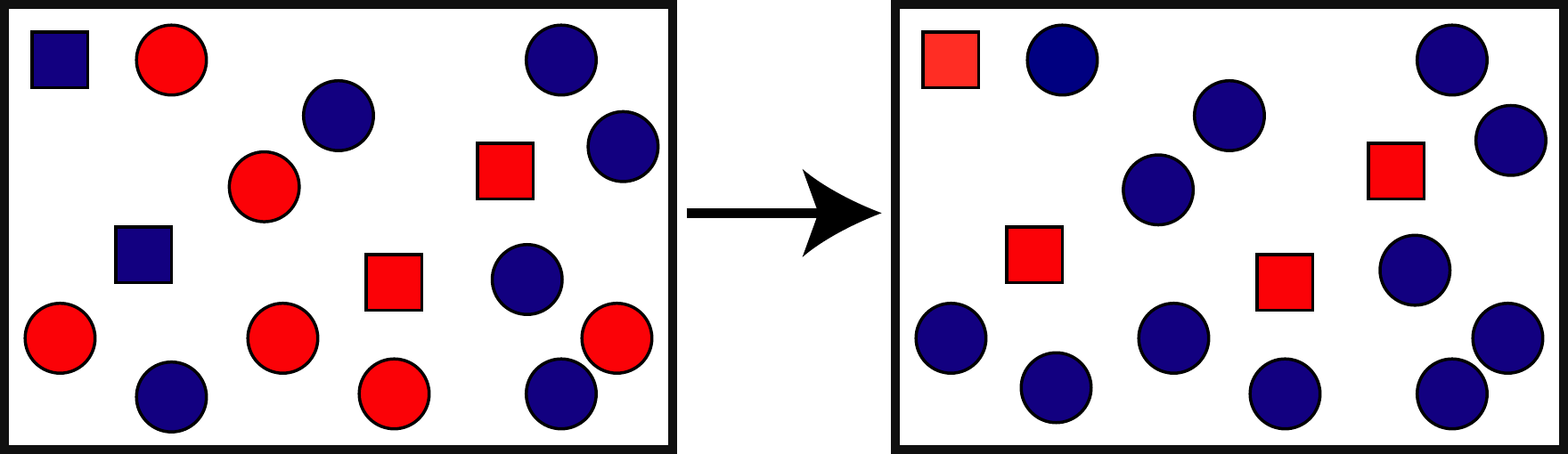}
\caption{Measurement in Szilard's second engine: Particle type variable
	$\PTyp\in\{\TSquare,\TCircle\}$ is depicted by shape and memory variable
	$\PMem\in\{\StZero,\StOne\}$ by color. (Left) initially uncorrelated
	demon-particle states---particle type is not correlated with memory (shape
	is not correlated with color). (Right) Configuration of the gas after
	measurement. Tracking from the left diagram to the right, the measurement
	process establishes a correlation between color ($\PMem$) and shape
	($\PTyp$): $\TSquare \to$ {\color{red}{red}} and $\TCircle \to$
	{\color{blue}{blue}}. There are only $\RSquare$s and $\BCircle$s.
	}
\label{fig:measure} 
\end{figure}

The first step of the control cycle is \emph{measurement}. Initially, the
ensemble's particle-type distribution is given by $\rho_0 (\PTyp)$ and the
distribution $f(\PMem)$ of the memory variable $\PMem$ is uncorrelated to particle
type: $\Pr(\PTyp,\PMem) = \rho_0(\PTyp) f(\PMem)$. We choose the parameter $\gamma$ to represent the initial distribution over the memory state of the particles, so that $f(\PMem)$ is initially distributed as $\Pr(\PMem_i = \StZero) = \gamma$ and $\Pr(\PMem_i = \StOne)=1-\gamma $.   During measurement, the current type $\PTyp_i$ of each particle
is imparted to its memory $\PMem_i$ such that each type $\TSquare$ ($\TCircle$)
particle has its $\PMem$ variable set to $\StZero$ ($\StOne$). Here, the
distribution $f(\PMem)$ changes so that the conditional distribution $f(\PMem_i|\PTyp_i)$ is deterministic or, equivalently,  the joint distribution over $\PTyp$ and $\PMem $ is given by nonzero elements $\Pr (\PMem_i = \StZero ,\PTyp_i = \TSquare ) = \delta$ and $\Pr (\PMem_i = \StOne ,\PTyp_i = \TCircle ) =1-\delta $ . See Fig. \ref{fig:measure},
where particle type is depicted via shape and particle memory via color.

Now, the cylinders slide relative to one another until the semi-permeable
membrane ends come into contact. In doing so, the semipermeable membranes
separate the particles by type. This is done without any input of work or heat
since, from the perspective of each particle, its container is merely being
translated; as demonstrated in Fig. \ref{fig:Control_SeparationSliding}. This
transformation separates the particles into one of two compartments. Particles
that were initially type $\TSquare$ are all in the left volume ($L$) bounded by
the pistons $\TSquare$ and $\TSquare'$; those that were type $\TCircle$ have
been shifted to the right container ($R$) that is bounded by the pistons
$\TCircle$ and $\TCircle'$. 

Time scales are important here. The separation must happen sufficiently slowly that the gas is always in equilibrium with respect to the container's spatial
volume, but fast enough that no particles transition types during the process.
This is not a generally prohibitive constraint, as we can assume the time-scale
for a gas to fill its container uniformly is generally quick. After the separation, each
particle type exists independently in a container of the same size as the
initial container. 

\begin{figure}
\includegraphics[width=\columnwidth]{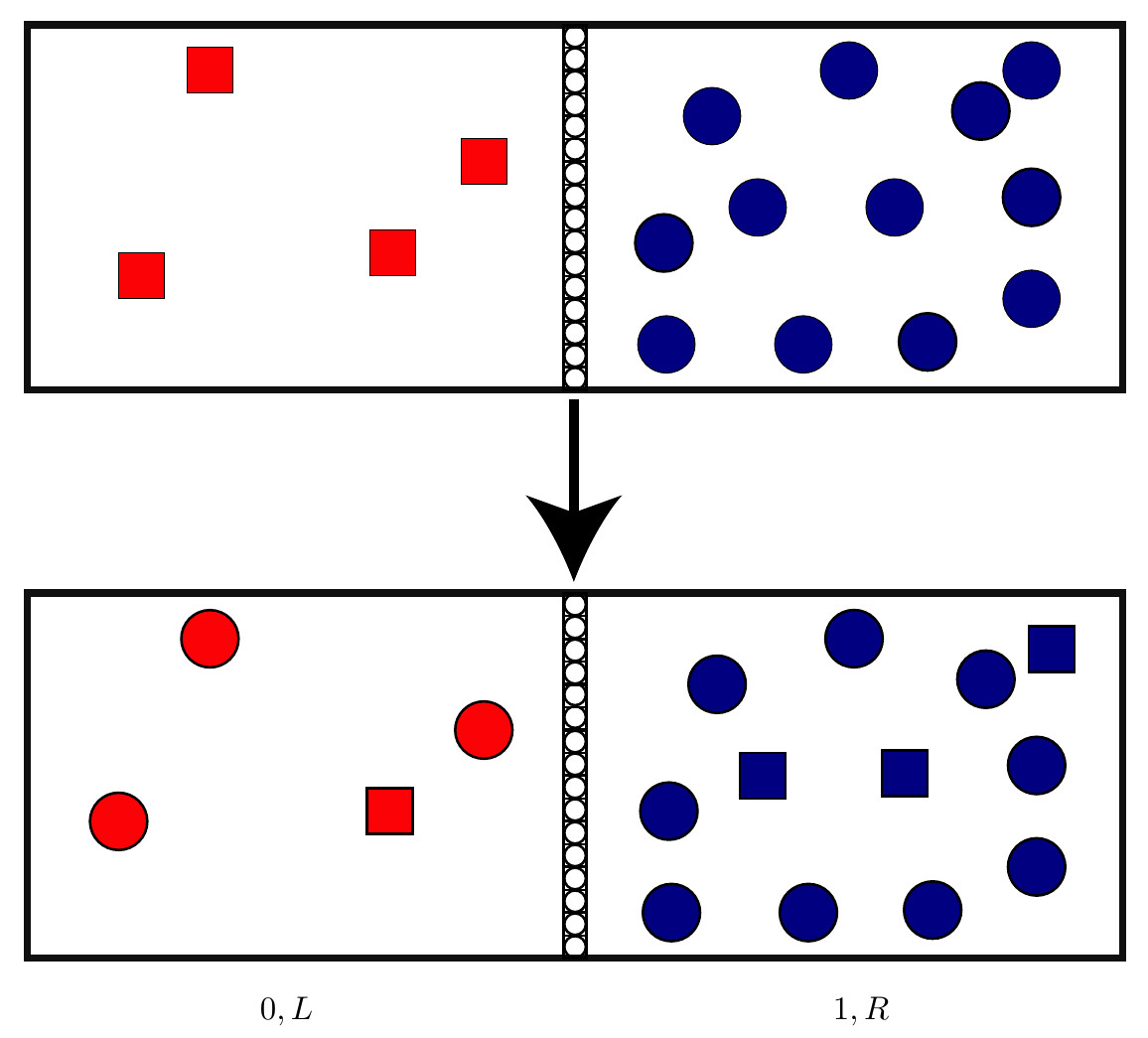}
\caption{Particle-type equilibration: (Top) Deterministic
	distributions $\Pr_L(\PTyp=\TSquare) = \Pr_L(\PMem=\StZero)=1$ and
	$\Pr_R(\PTyp=\TCircle) = \Pr_L(\PMem=\StOne)=1$ at the end of
	sliding-separation of Fig. \ref{fig:Control_SeparationSliding}. (Bottom)
	Distributions  after a period of particle-type conversion. Particles are no
	longer separated by shape type $\PTyp \in \{\TSquare,\TCircle\}$, but still
	by memory state (color) $y \in \{\StZero,\StOne\}$). That is, $\rho_L
	(\PTyp) = \rho_R (\PTyp) = \rho_0 (\PTyp) $ but the memory state
	distribution in each compartment remains deterministic.
	}
\label{fig:Control_Result_LongtermStates} 
\end{figure}

At this point, we introduce the subscript $L$ or $R$, to denote whether we are
discussing the distribution of particles in the left or the right compartment.
Each compartment is no longer in equilibrium with respect to the type variable.
See Fig.  \ref{fig:Control_Result_LongtermStates}(Top). In principle, we can
recover an equilibrium distribution with respect to $H_0$ individually within
the containers by waiting for the system to re-thermalize with the heat bath or
\emph{control} the process with a protocol involving the input or output of
work. See Fig. \ref{fig:Control_Result_LongtermStates}(Bottom). The latter is
discussed in detail in Sec. \ref{sec:secondengine} shortly.

\newcommand{\KE}{ \text{KE}_\text{avg} }

At this point in the engine's operation, Szilard claims the ``entropy has
certainly increased''. This is a familiar process and we expect it to increase
the system entropy, since we increase the gas' effective volume. The entropy
change $\Delta S$ from the initial macrostate to the final macrostate in which
the particles have re-achieved equilibrium can be found by the Sakur-Tetrode
equation (detailed in App. \ref{app:EntropyChange}), yielding:
\begin{align}
\frac{\Delta S}{N} & = -\kB  \left( \delta \ln  \delta
  + (1-\delta) \ln (1-\delta) \right) \nonumber \\
  & \equiv S(\delta) 
  ~.
\label{eq:EntropyChange}
\end{align}
The system's entropy increased, as Szilard claimed. If we re-establish the
equilibrium distribution reversibly (through a \emph{control} protocol) instead
of spontaneously, then there must be a corresponding decrease $-S(\delta)$ in
thermal reservoir entropy. Note that we cannot easily move the cylinders back
into each other now, since there are particles of both types on each side of
the semipermeable membranes. 

\begin{figure}
\includegraphics[width=\columnwidth]{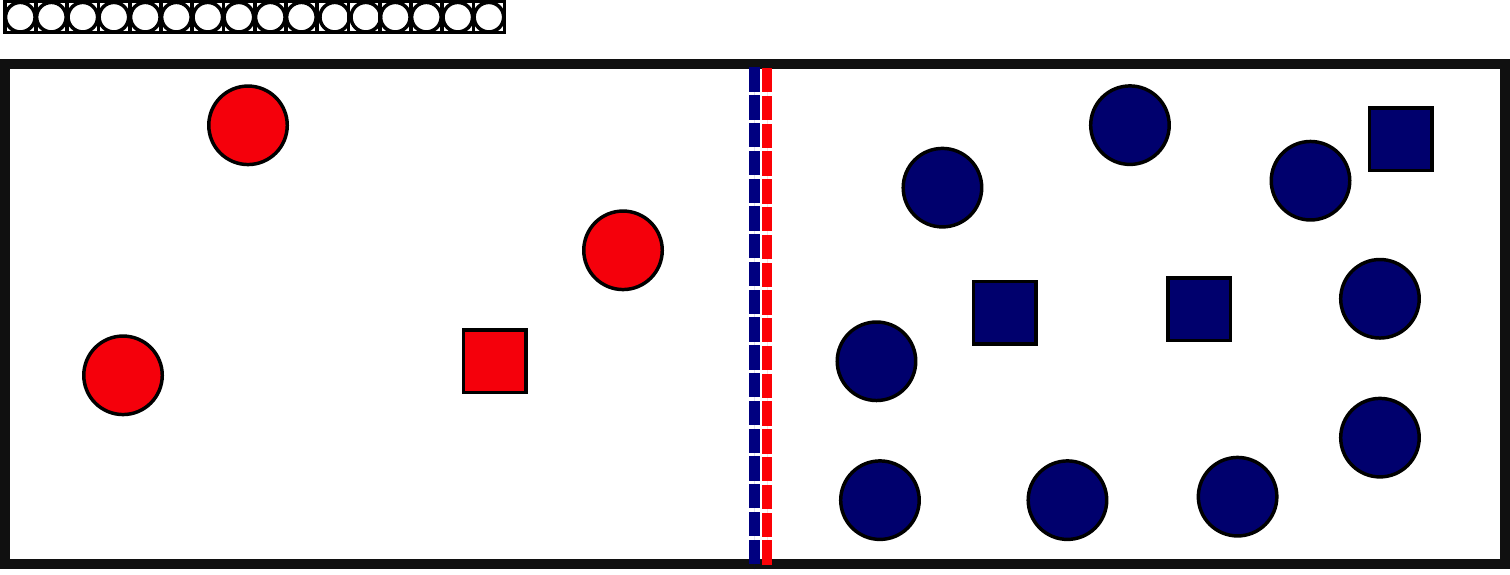}
\caption{First step to reintegrate molecules leading to the initial macrostate,
	replacing the type-semipermeable membranes with memory-state
	semipermeable membranes.
	}
\label{fig:TypeColorMembraneSwap} 
\end{figure}

Up to this point in the control protocol, the analysis only addressed the
thermodynamics of particle type (shape)---variable $\PTyp_i$---not particle memory
(color). That is, we have yet to use the particles' $\PMem_i$ variable. The engine
now makes clever use of its memories ($\PMem_i$) by exchanging the particle-type
semi-permeable membranes with membranes that are semipermeable to memory states
(color); see Fig. \ref{fig:TypeColorMembraneSwap}.

\begin{figure}
\includegraphics[width=\columnwidth]{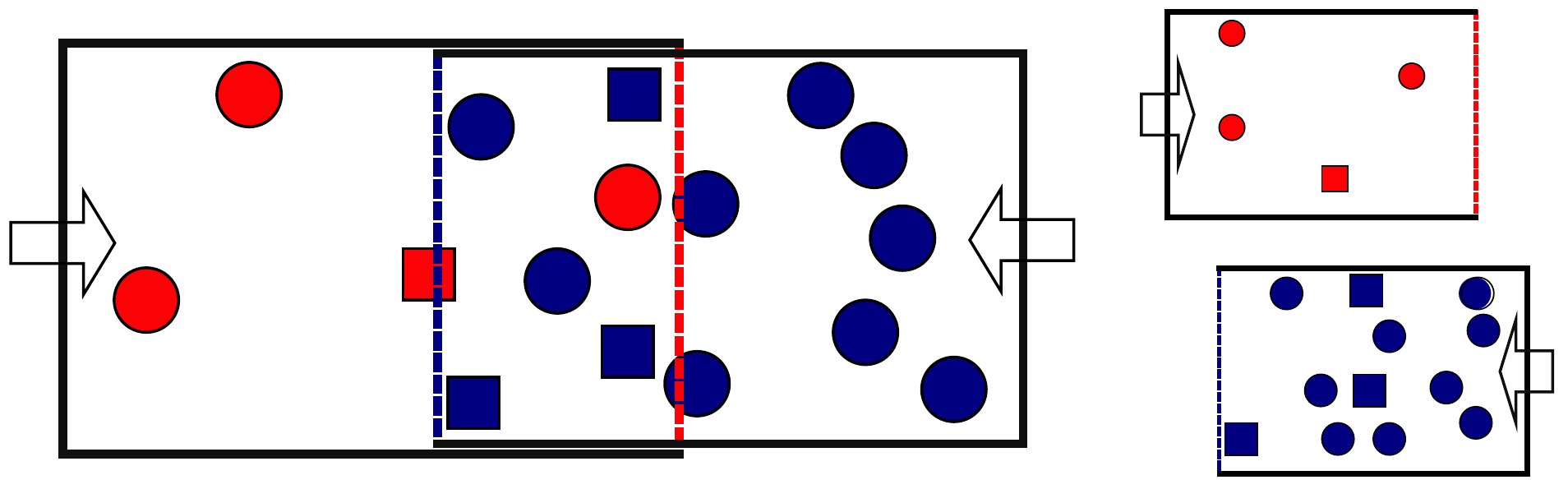}
\caption{Second step of reintegrating particles to return to the initial
	macrostate, leveraging the memory variable $y$ with the newly inserted
	memory-state (color) semipermeable membranes.
	}
\label{fig:ColorMembraneSlide} 
\end{figure}

The system is then ready to operate the reverse strategy as
when first type-separating the particles to bring them back into the same
volume. See Fig. \ref{fig:ColorMembraneSlide}. Again, this is accomplished
work-free, given that the process is done on the proper time scale. Now that the particles are back within the original volume (L) again, they are no longer separated by color or shape. In this \emph{erasure} process, we
manage to bring the system back to its initial $\rho_0(\PTyp)$ macrostate, without interacting
with the heat bath. The change in entropy for the system over the entire
protocol cycle is, then, zero. The thermal reservoir, however, had a net
decrease of entropy. At this point in the cycle, Szilard appeals to the
validity of the Second Law, stating that \cite{Szil29a}:
\begin{quote}
If we do not wish to admit that the Second Law has been violated, we must
conclude that $\ldots$ the measurement of $\PTyp$ by $\PMem$, must be accompanied by a
production of entropy.
\end{quote}
Szilard's associating measurement with a change in thermodynamic entropy and
giving the functional form Eq. (\ref{eq:EntropyChange}) of the latter
anticipates Shannon's communication theory and its measure of information
\cite{Shan48a} by nearly two decades.

\begin{figure}
\includegraphics[width=0.6\columnwidth]{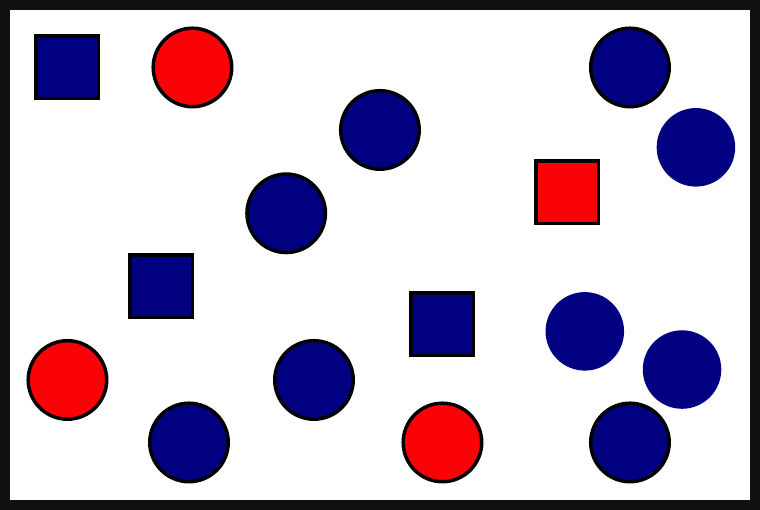}
\caption{Reintegration with sliding the memory-state (color) semipermeable
	membranes recovers the original distribution over particle type in the
	initial container.
	}
\label{fig:FinalColorTypeEquilibrium} 
\end{figure}

The careful reader will notice several issues that require further
investigation and definition. First, Szilard does not specify a mechanism that
stores the memory variable $\PMem$. Second, he does not investigate the work
required to drive the reversible control transformation he postulates. Third,
one notes that the final distribution over the memory variable $\PMem$, while not
correlated with type variable $\PTyp$ at the cycle's end, is necessarily
distributed so that $N \delta$ particles are in the $y = \StZero$ memory
state and $N(1-\delta)$ particles are in the $y = \StOne$ memory state; that is,
unless we include an additional erasure step that resets $y$ to some arbitrary
initial distribution. In addressing these (and related) concerns we shall see
that, while the selection of the initial distribution $f(\PMem)$ over memory
variables $\PMem_i$ is arbitrary, the choice impacts the thermodynamic costs of
measurement and erasure. First, we investigate the bounds on the work required
to perform Szilard's reversible control transformation.

\section{Engine Version 2.5}
\label{sec:secondengine}

During the control step, each compartment begins in a nonequilibrium
(completely deterministic) macrostate $\rho_L(\PTyp)$ (or $\rho_R(\PTyp)$)
(Fig.  \ref{fig:Control_Result_LongtermStates} (Top)) and ends in the canonical
equilibrium macrostate $\rho_0$  (Fig. \ref{fig:Control_Result_LongtermStates}
(Bottom)). To understand the effects of this transformation, we appeal to
recent developments in information theory and stochastic thermodynamics
\cite{Cover2012,Seifert2012,Parrondo2015} that allow us to connect the Gibbs
\emph{statistical entropy}:
\begin{align*}
S(\rho) & = -\kB \!\!\!\! \sum_{ \PTyp \in \{\TCircle,\TSquare\} }
  \!\!\!\! \rho(\PTyp) \ln \rho(\PTyp) \\
  & = \kB \langle - \ln \rho \rangle_\rho \\
\end{align*}
to the energetics of the isothermal equilibration process.

The two compartments ($L$ and $R$) interact separately with the heat bath, so
we take the following process to be executed independently within each
compartment. As such, we drop the $L$ and $R$ subscripts for clarity and take
the final extensive quantities to be of the form $S(\rho) \equiv S(\rho_L) + S
(\rho_R)$. Moreover, since the memory state remains deterministic within each
compartment throughout the control process, the only relevant distribution is
the marginal distribution---$\rho_L(\PTyp)$ or $\rho_R(\PTyp)$---over
particle type.

Assuming perfect control of the Hamiltonian at any point during the
transformation allows us to design the most efficient protocol for the
equilibration process. Consider the particle-cylinder system immediately after
particle separation, in contact with a thermal bath at temperature $T$.
Initially, the Hamiltonian is that given in Eq. (\ref{eq:Hamiltonian}). We
break the process into two distinct steps, both steps are executed within each
compartment as follows. First, we instantaneously shift the Hamiltonian from
${H}_0$ to $H_\rho= -\kB T \ln \rho $. Tautologically, $\rho$ is now
the equilibrium distribution since by definition the canonical equilibrium
probability distribution is $\rho = e^{-\beta H_\rho}$. Shifting the
Hamiltonian requires a minimum amount $W_{\Delta H}$ of work given by the
difference $\Delta H$ in the system's total energy under the two Hamiltonians:
\begin{align*}
W_{\Delta H} = \langle H_\rho\rangle_\rho - \langle H_0 \rangle_\rho
  ~.
\end{align*}
Next, we quasistatically shift the Hamiltonian back to $H_0$, which keeps
the system in equilibrium by definition.

\newcommand{\Ures}{ U_\text{res} }
\newcommand{\Sres}{ S_\text{res} }
\newcommand{\Qres}{ Q_\text{res} }
\newcommand{\Usys}{ U_\text{sys} }
\newcommand{\Ssys}{ S_\text{sys} } 
\newcommand{\Fsys}{ F_\text{sys} } 
\newcommand{\Wqs}{ W_\text{qs} }
\newcommand{\Stot}{ S_\text{tot} } 
\newcommand{\Wdrive}{ W_\text{drive} } 

The transformation is now complete---the Hamiltonian returned to $H_0$ and
the system's macrostate is given by $\rho_0$. Energy conservation in the second step implies that
thermal reservoir and system energies change according to the work $\Wqs$
invested in the transformation:
\begin{align*}
 \Wqs= \Delta \Ures + \Delta \Usys 
  ~.
\end{align*}
Assuming the reservoir maintains constant volume, we write the $\Wqs$ in
terms of initial and final free energies:
\begin{align*}
\Wqs = F(\rho_0) - F(\rho)
  ~.
\end{align*}
(Appendix \ref{app:FreeEnergies} gives the details.) Then, the total work
$\Wdrive \equiv \Wqs + W_{\Delta E}$ to drive the two-step transformation is:
\begin{align*}
\Wdrive = \langle H_0 \rangle_{\rho_0} - \langle H_0\rangle_\rho
  + T S(\rho) - T S(\rho_0)
  ~.
\end{align*}
Each term is readily interpreted in the present setting.

When considering the sum of both compartments---recall $\langle H_0
\rangle_{\rho} = \langle H_0 \rangle_{\rho_L} + \langle H_0
\rangle_{\rho_R}$---the energy expectation values for $\rho$ and $\rho_0$ are
the same. The average kinetic energy $KE_{avg}$ will be the same since the
whole system is thermalized to the same temperature, so we can neglect its
contribution. For the initial nonequilibrium distribution $\rho$ we have:
\begin{align*}
\langle H_0 \rangle_{\rho}  =
  N \epsilon_\SubTCircle \delta
  + N \epsilon_\TSquare (1-\delta)
  ~.
\end{align*}
And, under the $\rho_0$ distribution:
\begin{align*}
\langle H_0 \rangle_{\rho_0}
   = &N \delta
  ( \epsilon_\SubTCircle \delta + \epsilon_\TSquare (1-\delta))
  \\
  & + N (1-\delta)
  ( \epsilon_\SubTCircle \delta + \epsilon_\TSquare(1-\delta))
 ~.
\end{align*}
$\langle H_0 \rangle_{\rho_0}$ simplifies trivially to $\langle H_0
\rangle_{\rho}$. Together they make no contribution to $\Wdrive$. The
$TS(\rho)$ term vanishes since the initial distribution of particle types
within each compartment ($L$ and $R$) is deterministic. The final term, the
equilibrium state entropy, is $S(\rho_0) = NS(\delta)$. And so:
\begin{align*}
W_{drive} & = -TS(\rho_0) \\
  & =   -NTS(\delta)
  ~.
\end{align*}

It is now clear that the thermodynamic cycle is an engine. The driving work is
negative, signifying that there is an opportunity to extract work from the heat
bath. Once again, we are faced with the reality that the process of measurement
must contain compensating thermodynamic costs or admit that Szilard's second
engine is a type of perpetual motion machine.

\begin{figure}
\includegraphics[width=\columnwidth]{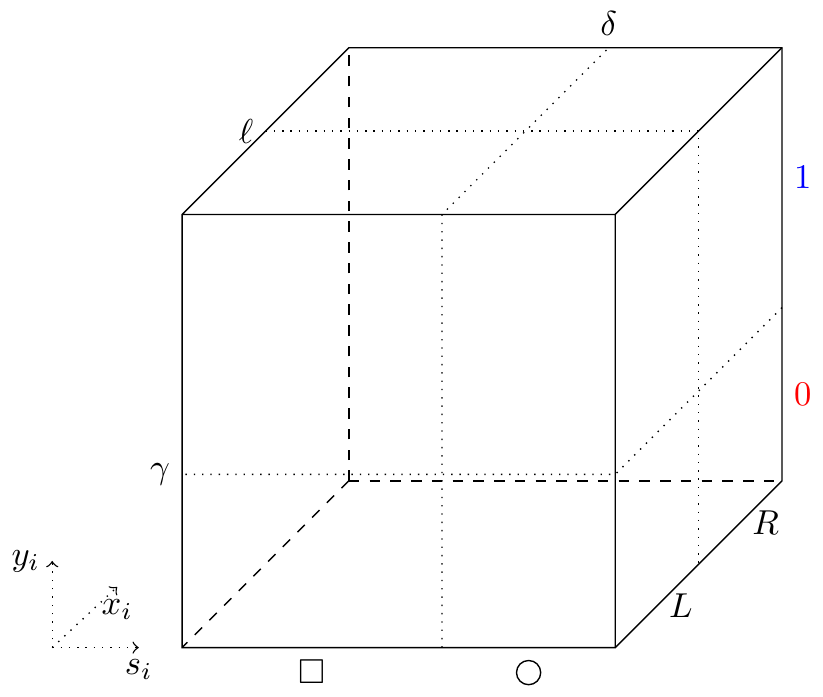}
\caption{Markov partition of a demon-particle's state space---3D unit box. The
	$i^{\text{th}}$ particle's position on the $\PTyp$ axis corresponds to its
	particle type as $\TypRV_i=\TSquare$ when $\PTyp_i < \delta$ (and
	$\TypRV_i = \TSquare$ when $\PTyp_i > \delta$). The $\PMem$-axis partition
	corresponds to the memory state as $Y_i=\StZero$ when $\PMem_i < \gamma$
	(and $\MemRV_i = \StOne$ when $\PMem_i >\gamma$). The depth dimension,
	parametrized by $\ell$, corresponds similarly to particle position being in
	the left or right compartment. Note that $\ell$ must be equal to
	$\frac{1}{2}$ for the $L \leftrightarrow R$ transition to always be work
	free, as Szilard noted.
	}
\label{fig:StateSpace} 
\end{figure}

\begin{figure*}
\includegraphics[width=1.8\columnwidth]{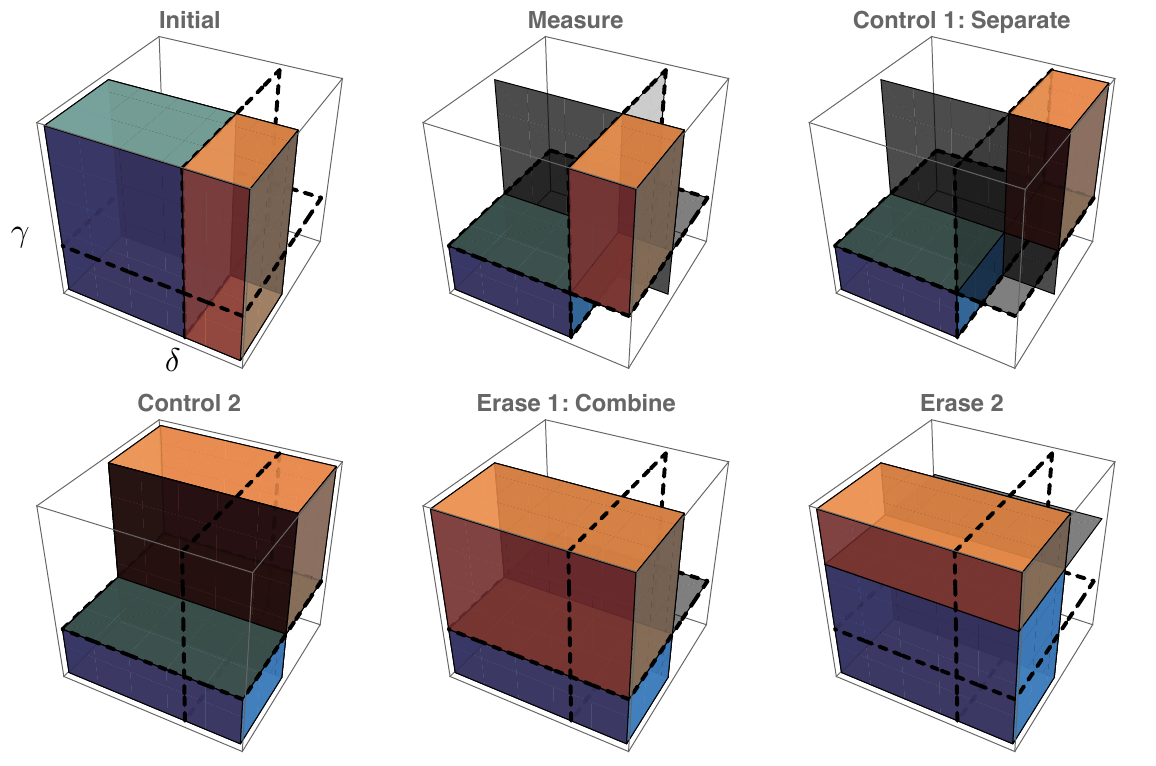}
\caption{Second Szilard engine's action on the unit-cube demon-particle state
	space decomposed into individual steps on an initially uniform
	distribution. Color illustrates which particles start as which type. The
	ideal barriers that are used to execute the protocol are depicted as
	dash-outlined gray planar partitions.}
\label{fig:SzilardMappings} 
\end{figure*}

\section{Demon Gas as a Thermodynamical System}
\label{sec:SzilardMap}

To investigate the cost of measurement thermodynamically, we must choose a
specific implementation of the device. We start with a 3-dimensional unit cube
in contact with a heat bath, inside are $N$ particles moving about thermally.
The previous section established that the work extracted by Szilard's engine is
independent of the energy difference $\Delta \epsilon$. We are, then, free to
set this difference to zero---yielding a box of particles that are all
identical according to $H_0$. The particles need not interact with each other
to perform any of the necessary operations, so we can choose them to be
noninteracting. Thus, our system is an ideal gas of $N$ identical particles.

The membranes separating the particles into the $L$ and $R$ compartments slide
along the box's $\PPos$ axis. We take all particles to start in the region
$\PPos < \ell $, with an ideal barrier inserted along $\PPos = \ell$ to keep
the particles from moving thermally into the region $\PPos>\ell$. In this way,
we defined two compartments $L$ and $R$ corresponding respectively to
$\PPos_i<\ell$ and $\PPos_i>\ell$.

We still need an operational definition of particle type which satisfies
Szilard's requirements that there is a fixed particle-type equilibrium and that
particles convert monomolecularly ($N$ is constant) from one type to another.
For our particle type, we choose the position of a particle along the $\PTyp$
dimension. If the coordinate of a particle is $\PTyp_i<\delta$ or
$\PTyp_i>\delta$ we consider it to be particle type $\TSquare$ or $\TCircle$,
respectively. As the particles move about thermally, they cross back and forth across the line
$\PTyp = \delta$ which exactly models Szilard's monomolecular conversion.
Additionally, by choosing the parameter $\delta$ we are able to set our
equilibrium distribution over particle type using the gas' tendency to quickly
fill its container uniformly.

This choice for particle type also allows us to define semipermeable
particle-type membranes as ideally impermeable membranes that cover only the
region associated with the relevant particle type. It is well known that the
physical position of particles stores information \cite{Landauer1961}. And so,
we choose a particle's memory states to be stored in its $\PMem$ coordinate
with $\StZero$ ($\StOne$) corresponding to $\PMem_i < \gamma$ ($\PMem_i >
\gamma$). See Figure \ref{fig:StateSpace} for the full partitioning of the
demon-particle system.

We are now ready to (i) analyze this engine's thermodynamics, (ii) set up the
symbolic dynamics for the gas particles, and (iii) analyze the engine's
intrinsic computation.

\subsection{Thermodynamics}
\label{sec:Thermodynamics}

The model introduced in Ref. \cite{Boyd14b} allows us to easily probe the
thermodynamics of each step in the Szilard Engine V. 2.5 operation, as just
described in Sec. \ref{sec:SecondEngine}. Given this representation of
Szilard's second engine, the overall thermodynamic cycle is the series of
transformations shown in Fig. \ref{fig:SzilardMappings}: \emph{measure},
\emph{control}, and \emph{erase}. These operations are executed by inserting,
sliding, and removing barriers.

The \emph{measure} step, for example, involves three barriers. First, we insert
a barrier along $\PTyp=\delta$. This is thermodynamically free, since the gas
is identical on either side of the barrier. Next, we use a barrier
perpendicular to the $\PMem$ axis that extends until $\PTyp<\delta$ to compress
the particles that are in the $\TSquare$ partition to fit entirely within the
$\StZero$ partition. Similarly, we use a barrier perpendicular to the $\PMem$
axis that covers $\PTyp>\delta$ to compress the particles that are in the
$\TCircle$ partition to fit entirely within the $\StOne$ partition. This establishes the
necessary correlation between type and memory state: all particles are either
$\RSquare$ or $\BCircle$.

For the \emph{control} step, the first operation separates particles by
type into either the $L$ or $R$ partition. This involves translating the
$\BCircle$ particles to the $R$ partition by inserting a barrier perpendicular
to the $\PPos$ axis at $\PPos=0$ that covers from $\PTyp= \delta$ to $\PTyp=1$.
Then, along with the $\PTyp>\delta$ section of the initial barrier, this barrier
translates the gas to the rear partition. This requires no interaction
with the heat bath, since the volume of the $\BCircle$ gas remains constant.

The second part of the control step expands along the particle-type dimension
by allowing the two sections of the particle-type partition corresponding to
$\PPos<\ell$ and $\PPos>\ell$ to slide independently of one another. The work
$W_\text{drive}$ the gas exerts on the barrier for an isothermal operation is calculated
easily as $- \int P \dd V$, with $P = N \kB T/ V$: 
\begin{align*}
W_\text{drive}
  & = -\int_{\ell\delta\gamma}^{\ell\gamma} \frac{N \kB T}{V} \dd V
  - \int_{\ell (1-\delta)(1-\gamma)}^{\ell(1-\gamma)} \frac{N \kB T}{V} \dd V \\
  & = N \kB T\left( \ln \delta + \ln (1-\delta)\right) \\
  & = -NTS(\delta)
  ~.
\end{align*}
This accords with the value calculated above. Thus, the model achieves the
ideal efficiency bound.

We can also calculate the thermodynamic costs of the measurement and erasure
transformations. In these, the gas' internal energy remains fixed and so
$Q_{sys}=-W_{sys}$. To investigate the energy that is dissipated in the heat
bath, we draw a relation between $Q_{sys}$, which is positive when heat flows
into the system from the bath, and $Q_{diss} = -Q_{sys}$, which is positive
when heat is being dissipated into the heat bath. For the measurement process,
we have:
\begin{align*}
Q_{M} & \! = \! -\! \int_{\ell \delta}^{\ell\delta\gamma}
  \frac{N\delta \kB T}{V} \dd V
  \! -\! \int_{\ell (1-\delta)}^{\ell(1-\delta)(1-\gamma)} \frac{N(1-\delta) \kB T}{V} \dd V \\
  & = N \kB T\left( -\delta \ln \gamma - (1-\delta) \ln (1-\gamma) \right)  \\
  & = N \kB T\left( \delta \ln \frac{1-\gamma}{\gamma}  - \ln(1-\gamma) \right) 
  ~.
\end{align*}

\begin{figure}
\includegraphics[width=\columnwidth]{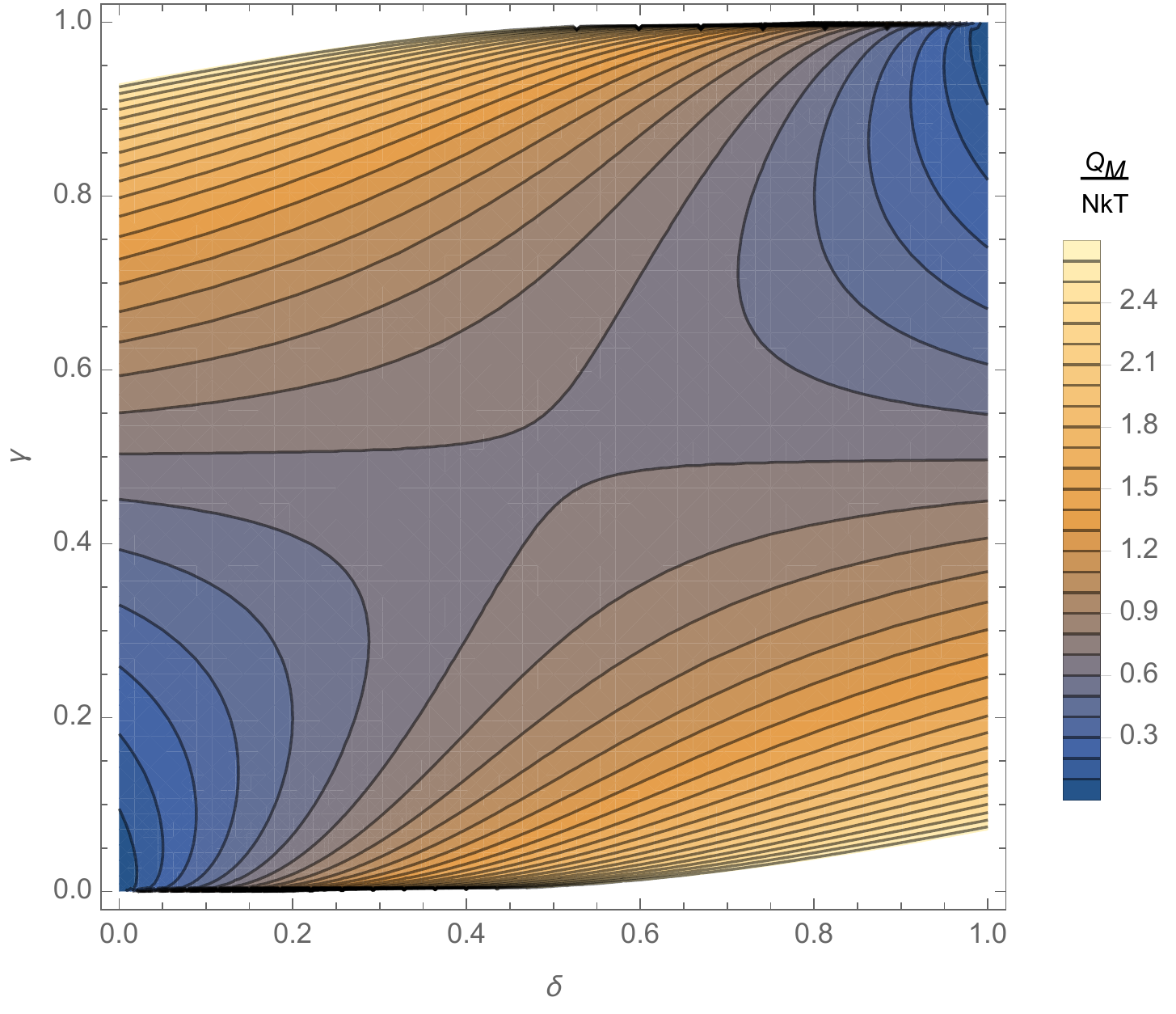} \\
\includegraphics[width=\columnwidth]{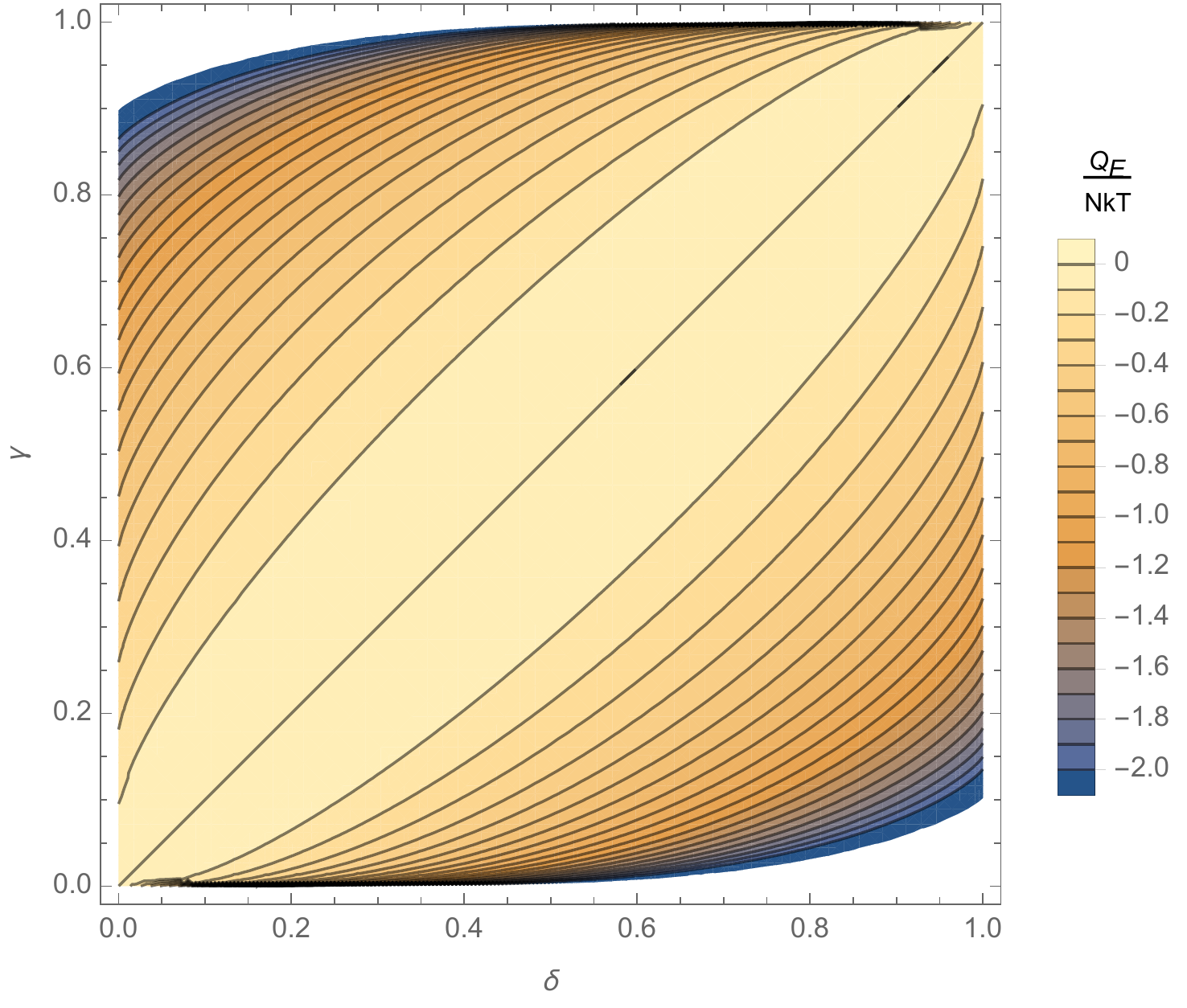}
\caption{Thermodynamic cost (heat dissipation) of measurement
	$Q_M/ N \kB T$ (Top) and erasure $Q_E/ N \kB T$ (Bottom) as a function of
	partition location parameters $\delta$ and $\gamma$.
	}
\label{fig:HeatMeasurementErasure} 
\end{figure}

Figure \ref{fig:HeatMeasurementErasure} (Top) displays a contour plot of the
measurement heat $Q_{M}$ as a function of the partition parameters $\gamma$ and
$\delta$. We see that the measurement thermodynamics strongly depends on these
parameters and that heat will always be dissipated during measurement. To
implement an efficient engine, then, we would select a set of parameters that
minimizes the heat dissipated in measurement.

Measurement is only part of the overall engine cycle, though. There is also the
erasure transformation. The first erasure step in Fig.
\ref{fig:SzilardMappings} translates the particles back into the same $\{L,R\}$
partition; similar to the first \emph{control} operation. The current model
makes it abundantly clear that this not sufficient to return the gas to its
initial state, though. The gas above and below the memory-state partition
(inserted at the beginning of the \emph{measurement} step) will not generally
have the same pressure. We require an additional step to return the gas to it's
initial maximum entropy state. Translating the boxes back to the $L$
compartment does not require any thermodynamic input or output so this final
step is the source of the thermodynamics of the erasure. The final step allows
the gas to slide the partition that separates our memory states until the
pressure on each side equalizes---until it rests at $\PMem=\delta$. The barrier
may then be removed at no cost or it may be left in the box and allowed to move
freely along with the next cycles without affecting the thermodynamics.
Calculating the energetic cost of this transformation is as simple as
preceding, yielding:
\begin{align*}
Q_{E} = N \kB T \left( (1-\delta)
  \ln \frac{1-\gamma}{1-\delta} + \delta \ln \frac{\gamma}{\delta}\right)
  ~.
\end{align*}

It is not surprising that the entropy cost of erasure vanishes when
$\delta=\gamma$, since then the barrier at $\gamma$ is already in the equal
pressure position before the final step. Figure
\ref{fig:HeatMeasurementErasure}(Bottom) shows that erasure does not incur a
cost: instead, the erasure provides yet another opportunity to extract energy
from the heat bath. This is as expected, as the erasure process always
increases the entropy of the system. However, examining $Q_M + Q_E$ we see that
choosing the parameters to maximize the energy extraction in erasure increases
the cost of measurement commensurately. Suggestively, the total thermodynamic
cost of measurement and erasure is algebraically independent of the parameter $\gamma$:
\begin{align*}
\frac{Q_M+Q_E}{N \kB T} & =
  \left( (1-\delta) \ln \frac{1-\gamma}{1-\delta} + \delta \ln
  \frac{\gamma}{\delta}\right) \\
  & \qquad + \left( \delta \ln \frac{1-\gamma}{\gamma}  - \ln(1-\gamma) \right) \\
  & = - (1-\delta) \ln (1-\delta) - \delta \ln \delta \\
\end{align*}
Or:
\begin{align*}
\frac{Q_M+Q_E}{NT} & = S(\delta)
  ~. 
\end{align*}
That is, the total combined cost of measurement and erasure depends only on
$\delta$, as in $NT S(\delta)$. This is exactly the energy necessary to
compensate for the work extracted from the heat bath during control. Since the
choice of $\gamma$ does not affect the total work extracted from the heat bat,
nor the total cost of the \emph{measurement} and \emph{erasure} processes
together, one can set $\gamma=\delta$ so that erasure is cost neutral and all
of the extracted work comes from the control process.

In this way, we need only consider the ``cost'' of measurement and the
``revenue'' from control. Of course, there is no net profit. Even in the most
efficient system, the Second Law holds. And, this was one of Szilard's main
points---the point that resolved Maxwell's paradox. By giving the demon (or
control subsystem) a physical embodiment and properly accounting for its
thermodynamics, there is no Maxwell demon paradox.

It is interesting to note that, just as in Ref. \cite{Boyd14b}, the distinction
between measurement and erasure turns out to be, in a sense, arbitrary. We may
increase or decrease the cost of one, but we do so at the expense of the other.
This harkens back to Szilard's original work, where he assigned entropy
production to the measurement and then goes on to demonstrate with a specific
measurement apparatus that the erasure step increases the entropy. (This
apparatus is discussed in Sec. \ref{sec:ThirdModel} below.) Szilard was not as
much concerned about when the entropy was produced, as that the production had
to be associated with the process of establishing and destroying correlation
between particle type and the  memory state.

This contrasts with the view advocated by Landauer and Bennett half a century
after---the logical irreversibility of erasure solely determines thermodynamic
costs \cite{Benn82,Land61a}. We now see, as others have recently emphasized
\cite{Shiz95a,Fahn96a,Bark06a,Saga12a,Boyd14b}, a more balanced view that there
is a generalized principle bounding the total costs of measurement and erasure.
Presumably, this is a constructive result that may lead to a design flexibility
in future information engine implementations.

\subsection{Computational Mechanics of the Gas Symbolic Dynamics}
\label{sec:GasSymbolicDynamics}

The course-graining of the microstate-space's unit box, depicted in Fig.
\ref{fig:StateSpace} is a Markov partition \cite{Laso85a} of the microstate
dynamics under the macroscopic thermodynamic transformations that make up the
Szilard engine. This immediately suggests defining a vector of binary variables
$( \TypRV_i \in \{\TSquare,\TCircle\},\MemRV_i\in \{\StZero,\StOne\},\PosRV_i
\in \{L,R\})$, sequences of which exactly track of the engine's microscopic
dynamical behavior.

At each protocol step a compound symbol $\TypRV\MemRV\PosRV$ is generated
according to the particle's location in the state-space box. For example, a
particle that ends a protocol step and generates the compound symbol $\TSquare
0 L$ corresponds to a particle that is currently type $\TSquare$, was particle
type $\TCircle$ when the most recent measurement was performed, and is in the
Left compartment.

We must remind ourselves that the state space of this gas is large. There are
$N$ particles in three dimensions, so the full state space represents a $3N$
dimensional dynamical system. However, since the particles are noninteracting,
Szilard's second engine is actually a collection (direct product) of $N$ 3D
particle state spaces. Applying computational mechanics' \emph{predictive
equivalence relation} collapses the $3N$-dimensional state-space to $3$
dimensions of equivalent causal states \cite{Crut12a}. Thus, we can use the
symbolic dynamics of a single particle to find the engine's effective
information processing behavior---and scale it to $N$ particles with a
pre-factor of $N$.

The problem simplifies even further since, having faithfully considered
Szilard's initial problem statement, it is clear that the $LR$ dimension of the
state-space box is redundant in the current model. It was useful in Szilard's
original construction to include a barrier that stops particles corresponding
to the different memory states from intermixing. However, the current engine
stores the memory and type states in positional coordinates, so the barrier
used to compress the gas in the cycle's \emph{measure} step already serves this
purpose. Thus, we do not even need the full $3$-dimensional state-space box to
model the system's information and thermodynamic action.

Instead, we examine the action of Szilard's second engine on a $2$-dimensional
projection onto the $\PTyp\PMem$ plane of the box in Fig.
\ref{fig:SzilardMappings}. The resulting $2$-dimensional map  is nearly
identical to the Szilard Map introduced in Ref. \cite{Boyd14b}, constructed by
considering Szilard's first or single-molecule engine. The only differences
between these maps is a different initial state distribution. In fact, we could
reconstruct the second Szilard engine to have the same initial state but, for
the purpose of illustration, we will investigate the map under the current
default memory state. At this point, though,  one fully expects the results
to agree with Ref. \cite{Boyd14b} in every fundamental sense.

We now track the probability density of a particle within the gas. Having
abandoned tracking each particle's exact position within the box by using the
course-graining into binary symbols, we now consider the actions of a
deterministic map on the probability density as a whole. Each step in the
process depicted in Fig. \ref{fig:SzilardMappings} compresses or expands the
probability density along a particular dimension. The composite map that
includes each step when $\delta = \gamma$ is given by:
\begin{align*}
\tau_{\text{Szilard}} (\PTyp,\PMem)
  = \begin{cases} 
      \left(\frac{\PTyp}{\delta}, \PMem \delta\right) & \PTyp<\delta  \\
      \left(\frac{\PTyp-\delta}{1-\delta}, \delta+\PMem(1-\delta)\right) & \PTyp>\delta 
   \end{cases}
  ~.
\end{align*}
Appendix \ref{app:DeterministicMaps} gives the maps for each individual step.

In accordance with the analysis in Ref. \cite{Boyd14b}, if we choose to leave
out the memory-state partitions that are added each cycle, we build up the same
self-similar interleaving within the particle's state-space probability
distribution as seen in the Baker's Map \cite{Stro94a}. While the probability
density is not uniform throughout each step of the map, we find that the
distribution over the state space is uniform and constant for the composite map
$\tau_{\text{Szilard}}$ above that includes each step in the protocol.

\subsection{Information and Intelligence}
\label{sec:InfoIntel}

\begin{figure}
\includegraphics[width=0.7\columnwidth]{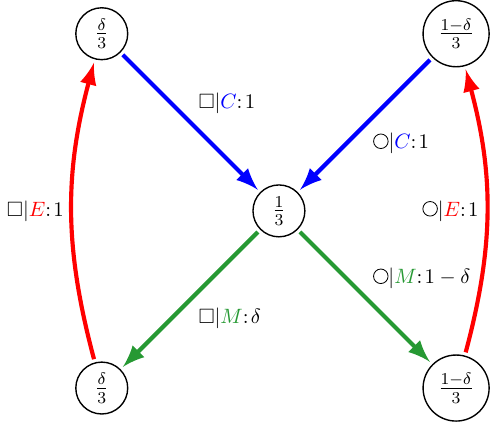}
\caption{\ET for the particle type ($x$) subsystem. Protocol steps are
	designated by color: (control, measure, erase) $\Leftrightarrow$ (blue,
	green, red). Numbers inside states correspond to the asymptotic state
	probability. The transition notation $s|d:p$ corresponds to emitting the
	symbol $s$ with probability $p$ given the driving symbol $d$.
	}
\label{fig:XMYM_type} 
\end{figure}

\begin{figure}
\includegraphics[width=0.7\columnwidth]{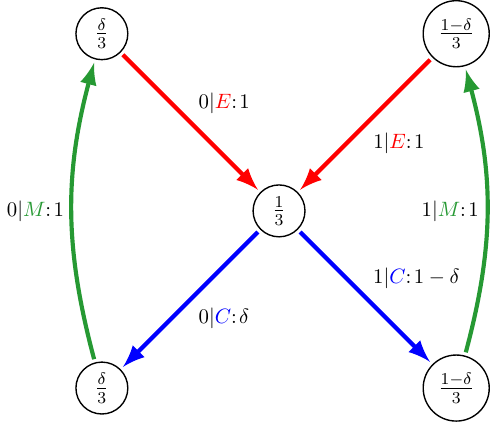}
\caption{\ET for the memory state (y) subsystem. Notation as in previous
	figure.
	}
\label{fig:XMYM_memory} 
\end{figure}

\begin{figure}
\includegraphics[width=\columnwidth]{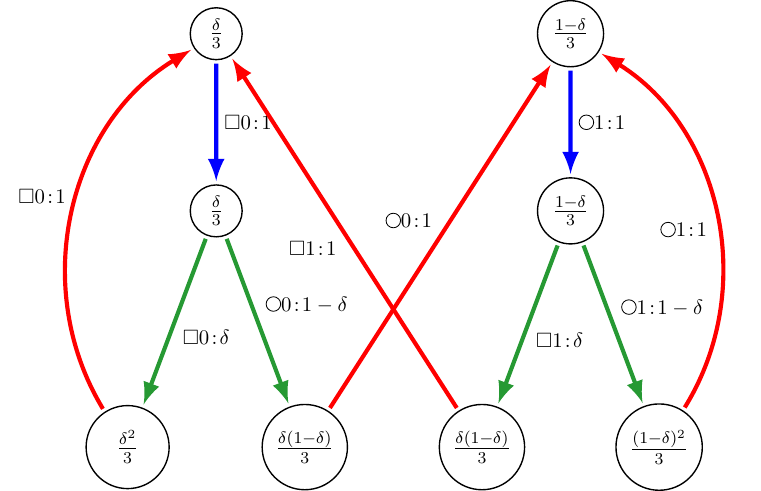}
\caption{\EM for the joint demon-particle system: Protocol steps designated by
	color as in Figs. \ref{fig:XMYM_type} and \ref{fig:XMYM_memory}. Driving
	symbols are suppressed in the transition notation for clarity---$s:p$
	corresponds to emitting the symbol $s$ with probability $p$.
	}
\label{fig:JM} 
\end{figure}

We again apply computational mechanics' predictive equivalence relation---now
not to the gas' microscopic state space but to the symbolic dynamics induced by
Markov partition of Fig. \ref{fig:StateSpace}. This leads directly to an \eT
\cite{Barn13a} that captures the information processing embedded in the
engine's operation. Figures \ref{fig:XMYM_type} and \ref{fig:XMYM_memory} show
the transducer for each dimension separately and Fig. \ref{fig:JM} shows the
\eM for the joint process. Then, retracing the steps in Ref. \cite{Boyd14b}
establishes that Szilard's first and second engines are informationally and
thermodynamically equivalent, though they arise from rather different
implementations.

Composing Figs. \ref{fig:XMYM_type} and \ref{fig:XMYM_memory} transducers with
the period-$3$ input process---that specifying the measure-control-erase
protocol, gives an \eM that generates the output process for particle type or
for memory state. (In this case, this is trivially implemented by dropping the
input symbols $\{M,C,E\}$ from the \eT transitions.)

The two resulting \eMs and that in Fig. \ref{fig:JM} are counifilar
\cite{Crut08b}. The processes are not cryptic and this greatly simplifies
calculating various informational properties. For example, the entropy rate of
the joint system's machine (Fig. \ref{fig:JM}) is $\frac{1}{3} \H(\delta)$ per
step, consistent with the analytical result for the Baker's Map from Pesin's
theorem. However, there is a slight variation from Ref. \cite{Boyd14b}'s
analysis of the statistical complexity $\Cmu$---the information in an \eM's
causal state distribution $\{\tbf{S} \}$. It is immediately clear from Figs.
\ref{fig:XMYM_type} and \ref{fig:XMYM_memory} that $\Cmu^x$ and $\Cmu^y$ are
equal. This was not the case in Ref. \cite{Boyd14b}. Thus, we see that the
different choice of initial state symmetrizes the stored information with
respect to particle type and memory state. The calculations for $\Cmu$ are
straightforward, nonetheless:
\begin{align*}
\Cmu^x & = \Cmu^y \\
   & = -\left[\frac{1}{3} \log_2 \frac{1}{3} + \frac{2}{3}\left(\delta \log_2 \frac{\delta}{3} +(1-\delta)\log_2 \frac{1-\delta}{3}\right)\right] \\
 & = \log_2 3 + \frac{2}{3} \H(\delta)
  ~.
\end{align*}
 
Similarly, we find that  $\Cmu^\text{joint} = \frac{4}{3} \H(\delta) + \log_2
3$. These are quantitatively different from the results in Ref.
\cite{Boyd14b}. However, that is the end of the differences. If we consider the
relationship between the three, we recover that $\Cmu^\text{joint} = \Cmu^x +
\Cmu^y - \log_2 3$. So, we see that the two engines have the same information
related to synchronization of their two subsystems. For the original
single-molecule engine these subsystems were demon memory and molecule
position; for engine Version 2.5, they are shape and color. In fact, we can
explicitly check that all other informational measures agree with Ref.
\cite{Boyd14b}. In doing so, we recover the asymptotic communication rate:
\begin{align*}
\lim_{L \to 0} \frac{\I[X_{0:L};Y_{0:L}]}{L} = \frac{1}{3} \H(\delta)
  ~,
\end{align*}
the correlation rate:
\begin{align*}
\lim_{L\to0} \frac{\I[S^x_{0:L};S^y_{0:L}]}{L} = \frac{1}{3} \H(\delta)
  ~,
\end{align*}
and the dependence of the correlation during the protocol steps,
which yields:
\begin{align*}
\I[X_0:Y_0|M] = \H(\delta)
  ~.
\end{align*}
The measurement step is where the single-symbol correlation is established. So,
it stands to reason that the correlation dependence is found to be entirely in
this step.

\section{Szilard's Third Engine: Measurement and strongly coupled systems}
\label{sec:ThirdModel}

The Szilard Map stores its memory state in an additional spatial dimension.
Section \ref{sec:SzilardMap} and Ref. \cite{Boyd14b} tease out the
thermodynamic and information-processing consequences of this choice. The
following introduces an alternative implementation of information storage---one
introduced by Szilard himself.

After concluding that the measurement process in his engines must generate
entropy, Szilard introduces a limit on the production of entropy from a binary
measurement:
\begin{align*}
e^{-S_\TSquare / \kB T} + e^{-S_\SubTCircle /\kB T} \leq 1
  ~,
\end{align*}
where $S_\TSquare$ and $S_\SubTCircle$ are the entropies that a protocol
generates when taking the measurement value $\TSquare$ or $\TCircle$,
respectively.
Investigating this limit further, he adopts a specific mechanical system that
performs the minimal measurement tasks that his engines require.

The essential tasks in measurement are as follows. First, establish a
correlation between the instantaneous value of a fluctuating variable $x$ and
another variable $y$. Second, store that value in the ``memory'' of the second
variable so that if $x$ changes, $y$ remains fixed. Finally, return to a default
state so that the system is ready to preform another measurement.

\begin{figure}
\includegraphics[width=\columnwidth]{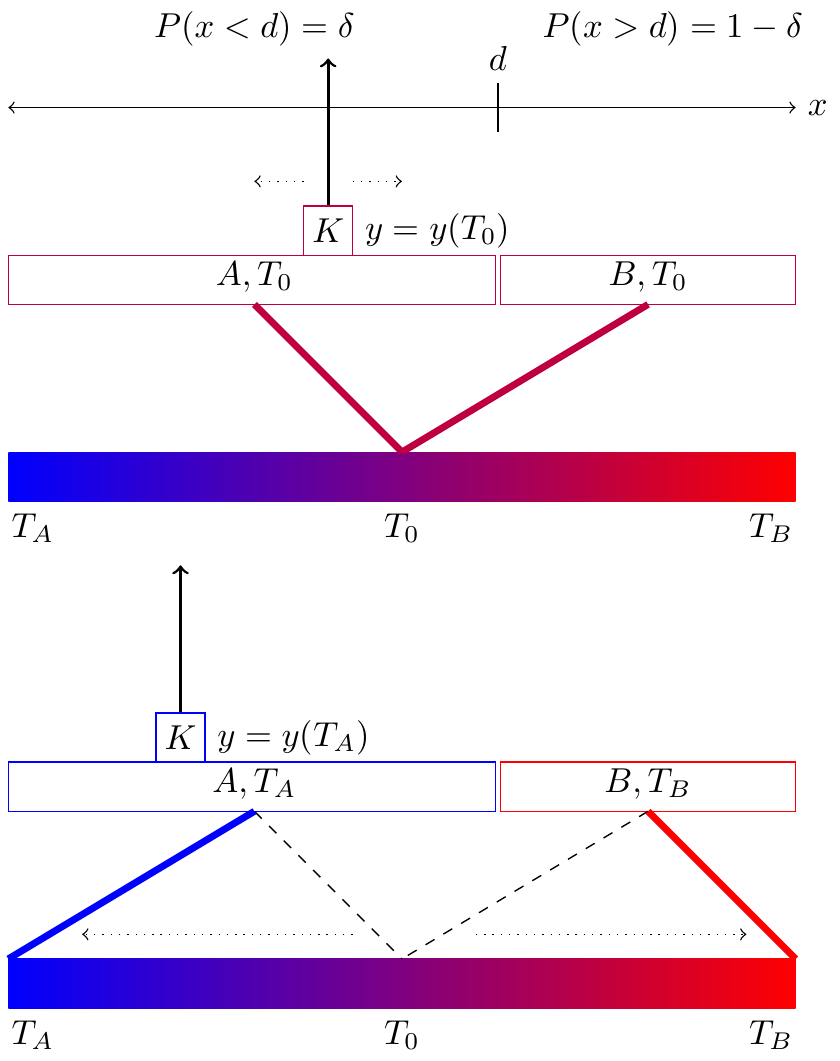}
\caption{(Top) Default state before measurement: Variable $x$ tracks the
	position of the pointer and $y$ is a function $y(T)$ of the temperature $T$
	of body $K$.
	(Bottom) Measuring position of the pointer by temperature of $K$. The
	pointer location at the time of measurement determines if $K$ is cooled
	to $T_A$ or heated to $T_B$. Consequently, $y$ is set to either $y(T_A)$
	or $y(T_B)$.
	}
\label{fig:third_construction_initial_measure}
\end{figure}

In this third construction of Szilard's, the variable to be measured $x$ is the
position of a pointer that moves back and forth according to a completely
general protocol, either stochastic or deterministic. The variable $y$ that
stores the position is a function of the temperature of a body $K$ that is
mechanically connected to the end of the pointer. As this pointer moves back
and forth, it brings $K$ in contact with one of two intermediate temperature
reservoirs, $A$ or $B$. These reservoirs are connected by movable
heat-conducting rods to a continuum of temperature reservoirs that span from a
cold temperature $T_A$ to a warm temperature $T_B > T_A$. Initially, both rods
are connected to an intermediate temperature $T_0$; see Fig.
\ref{fig:third_construction_initial_measure}. 

By coarse-graining the position of the pointer into two regions ($A,B$) we are
able to make a binary measurement. The measurement, which must happen during a
timescale for which the pointer is stationary, involves moving the connecting
rods through the continuum of heat reservoirs so that the intermediate
reservoir $A(B)$ is cooled(heated) to $T_A(T_B)$. In this process $K$ will
either become heated or cooled depending on where the pointer was, see Fig.
\ref{fig:third_construction_initial_measure}. This process can be done with
arbitrarily small dissipation, if the process is done slowly enough such that
the rod, the intermediate reservoirs, and $K$ remain in thermal equilibrium at
all times. We have now accomplished a binary measurement, where $K$ is either
at $T_A$ or $T_B$, depending on the position of the pointer at the moment of
measurement.

Next, the entire assembly of reservoirs is thermally isolated from the pointer
and $K$ so that, as the pointer continues to move, $K$ maintains its
temperature either at $T_A$ or at $T_B$, even as the pointer leaves the
interval it was in at the time of measurement. In this condition, the
measurement value is stored in $K$'s energy content. Now, to be ready to make
another measurement, the system must return to its initial state.  If one knew
with certainty $K$'s temperature, the system could be returned to the default
state without entropy cost: Simply wait until the pointer is in the region that
corresponds to $K$'s temperature, bring the system back into contact with the
reservoirs, and institute the measurement protocol in reverse. This is, of
course, actually two different protocols---and requires knowledge (measurement)
of the result of each measurement to decide which to implement on each cycle.

There is no single protocol that can blindly return the system to its original
state without producing entropy. Anticipating Landauer's well-known argument
for the bistable well \cite{Landauer1961} by more than three decades, Szilard
notes that an increase in entropy ``cannot possibly be avoided'' because
\cite{Szil29a}:
\begin{quote}
After the measurement we do not know \ldots whether [$K$] had been in
connection with $T_A$ or $T_B$ in the end. Therefore neither do we know whether
we should use intermediate temperatures between $T_A$ and $T_0$ or $T_0$ and
$T_B$.
\end{quote}

\begin{figure}
\includegraphics[width=\columnwidth]{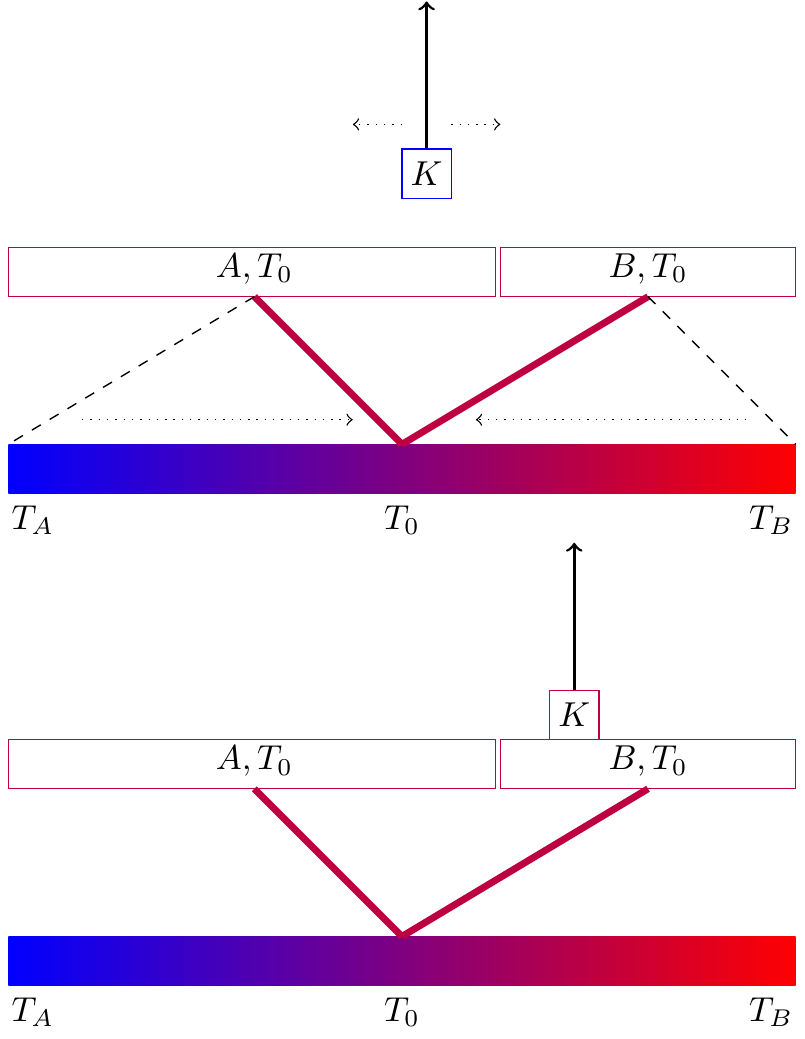}
\caption{(Top) While $K$ stores the location of the pointer at the time of
	measurement, $A$ and $B$ are returned to $T_0$. (Bottom) $K$ is returned
	to $T_0$ by thermal contact with $A$ or $B$, incurring an unavoidable
	entropic cost.
	}
\label{fig:third_construction_store_erase}
\end{figure}

We create, then, a single protocol that returns the system to its original
state---the ``erasure'' process---and measure its total entropy generation.
While the pointer is still uncoupled to the system, we return $A$ and $B$ to
the equilibrium temperature $T_0$. Once again, this can be done reversibly on
an appropriate timescale; see Fig. \ref{fig:third_construction_store_erase}.
Then, we bring $K$ back into thermal equilibrium. This step cannot be done
reversibly. This gives merit to the idea that erasure is the source of the
entropic cost. Quantitative accounting for the entropy generation reveals
additional insight.

All said, the body $K$ undergoes a cyclic process, so the net change in the
system entropy is zero. Thus, we consider the entropy change only in the
reservoirs. If the pointer was at a location that caused $K$ to cool (heat) to
$T_A$ $(T_B)$, then the reservoir's entropy increases (decreases) during the
measurement period by $\int \frac{dQ}{T} $. Similarly, when $K$ is returned to
$T_0$ the reservoir's entropy decreases (increases) by $\frac{\Delta E}{T_0}$.
We see that, while only the erasure process causes the entropy of the universe
to increase, both the measurement and erasure processes play a role in
increasing and decreasing the reservoir entropy. Szilard was unconcerned with
keeping measurement and erasure as two different actions since he already
concluded that it was possible for either to produce or consume the reservoirs'
entropic resources. This is an insight that only recently received renewed
attention \cite{Shiz95a,Fahn96a,Bark06a,Saga12a,Boyd14b}. Furthermore, Szilard
had also already concluded that the need to erase a binary random variable to a
default state had unavoidable entropic costs.  

To determine quantitatively the entropy gain from each measurement, Szilard
adopts a $2$-level system. The body $K$ can be on one of two energy levels: a
low energy state and a high energy state. Using standard canonical ensemble
calculations, he shows that in the limiting process where the probability of
the low (high) energy state at $T_A$  $(T_B)$ approaches unity the entropy
generated by each process is:
\begin{align*}
 S_A &=  -\kB \ln p \\
 S_B &= -\kB \ln q 
  ~,
\end{align*}
where $p=p(T_0)$ and $q=q(T_0)$ are the probabilities that $K$ is in the lower
and upper energy state at temperature $T_0$, respectively. Szilard ended his
analysis here. He does note that for this model:
\begin{align*}
e^{-S_A/k} + e^{-S_B/k} = 1
\end{align*}
and that this represents the minimum amount of entropy generation necessary
according to his limit:
\begin{align*}
e^{-S_A / \kB T} + e^{-S_B /\kB T} \leq 1
  ~. 
\end{align*}

\section{Szilard Measurement in Szilard Engines}

To complete our analysis of Szilard's arguments, we couple the Szilard
measurement device (SMD) above to one of his engines. For a simple physical
picture, we specialize to the more familiar single-particle Szilard engine
\cite{Szil29a,Benn82,Boyd14b}, where a classical particle in a box is used to
extract work from a temperature bath by inserting a partition and allowing the
particle to move the partition. In essence, this engine leverages the
measurement of a thermal fluctuation to do work. The considerations above show
that the multi-particle second engine has the same thermodynamic and
information processing behavior as the more-oft-quoted single-particle engine,
as treated in Ref. \cite{Boyd14b}. And so, we lose nothing by specializing to
his first, simpler model.

Now, take the SMD's pointer to be mechanically connected to the particle inside
the Szilard engine, so that the position of the pointer tracks the particle's
thermal motion. The SMD is calibrated so that the particle on the left-
(right-) hand side of the partition corresponds to the pointer being at
$x<\delta$ ($x>\delta$). The SMD is thermally isolated from the rest of the
engine, as having its own set of reservoirs is crucial to its operation. In
this example, the body $K$ plays the demon's role: the body changes length
depending on its energy state, allowing $K$'s state to select the engine's
protocol; for example, by the position of a switch connected to $K$.

In this way, the entropy generated by one engine cycle is the sum of the entropy
generated in the SMD's reservoirs and the particle-box system's reservoirs. The
particle moves thermally through the entire box, so the probability that it
falls in one or the other section of the box depends on the relative volume on
either side of the inserted partition. During a cycle, the entropy generation
in the system's reservoirs is proportional to either $\ln \delta$ or to $\ln
(1-\delta)$ depending on if the particle starts on one or the other side of the
partition, respectively. The mean entropy generated in the system reservoir
over many cycles is then:
\begin{align*}
\langle \Delta S_{res_s} \rangle
  & \propto \delta \ln \delta + (1-\delta)\ln (1-\delta) \\
  & = - H(\delta)
  ~.
\end{align*}

The mean entropy generation in the SMD's reservoirs over many cycles
of the measurement process is:
\begin{align*}
\langle \Delta S_{res_m} \rangle
  & \propto \delta S_A + (1-\delta) S_B \\
  & = - \delta \ln p - (1-\delta) \ln q
  ~.
\end{align*}
Adding the two contributions yields the average entropy generated in the
universe per cycle:
\begin{align*}
\Delta S
  & \propto (1-\delta) \ln \frac{1-\delta}{1-p} + \delta \ln \frac{\delta}{p}
  \\
  & = D_{KL}(\delta||\gamma)
  ~,
\end{align*}
where the relative information $D_{KL}(\cdot||\cdot)$ is positive for all
values of $\delta \neq p$ and vanishes when $\delta=p$.

We see that, once again, there is no Maxwell demon paradox, as the total
entropy generation is positive. In the case that $p = \delta$, the mean entropy
produced in the SMD reservoirs during the measurement cycle is exactly enough
to compensate the decrease of entropy in the system's reservoirs during the
work extraction. This encoding of the information yields the most efficient
cycle, in accordance with the analysis in the previous sections and Ref.
\cite{Boyd14b}.

However, there is no physical requirement that $p=\delta$. The inner workings
of body $K$ need not match where the partition between the physical regions $A$
and $B$ lies. It is tempting to make a direct comparison between $p$ and the
parameter $\gamma$ in Ref. \cite{Boyd14b}'s Szilard Map, but there is a
distinct difference between the two. Under the action of the Szilard Map, as
introduced in Ref. \cite{Boyd14b}, there is inherent interaction between the
parameters $\delta$ and $\gamma$ that manifests itself in the density of the
ideal gas that serves as the engine's ``working fluid''. When one part of the
gas is compressed into a particular memory partition, the size of that
partition determines the cost of the next step. If the partition is small, it
is ``more difficult'' to squeeze the same number of particles in. Consequently,
both $\delta$ and $\gamma$ appear in the cost of both measurement and erasure.
This coupling between the two dimensions becomes relevant when we take into
account the total cost of measurement and erasure, finding that $\gamma$ drops
out of the consideration. The result is an engine that is ideally efficient for
every parameter setting.

When coupling the SMD to the first engine, the importance in the inherent
interaction of $\delta$ and $\gamma$ in the Szilard Map becomes even more
apparent. Unlike $\gamma$, the engine is no longer ideally efficient for any
choice of parameter $p$. Instead, we must choose the distribution of $K$'s
energy states at the equilibrium temperature to have the same distribution as
the particles position states. If not, the engine suffers additional
dissipation from a mismatch of our system and our measurement device.
(Reference \cite{kolchinsky2017} recently considered the energy costs of such
mismatches.)

Looking across the sweep of progress since his original results, we now see
that Szilard's construction is a concrete example of Ashby's \emph{Principle of
Requisite Variety} \cite{Ashb57a}: the variety of actions available to a
control system must match the variety of perturbations it is able to
compensate. Specifically, Szilard recognized that a minimal control system for
a binary measurement must have two states. (Yet again, Szilard predated the
cybernetics era by several decades.)

However, we also see a stricter requirement used to avoid unnecessary
dissipation. The actual distribution over the controller's internal states
must be the same as the system's. This also touches on the general arguments
put forth in Ref. \cite{Boyd16d} that consider the efficiency of a
thermodynamically embedded \emph{information ratchet} which interacts with an
information reservoir to extract work. Finally, one sees a clear parallel
between the information-theoretic concept of optimal encoding \cite{Cover2012}
in which minimizing the memory needed to store a particular message, the
highest probability events are given the shortest codewords. In short, it
should not be surprising that it is optimal to match our controller to the
system. However, it is gratifying to see such a clear and straightforward
example---an example unfortunately ignored by Szilard's future colleagues.

The SMD model of measurement also provides a clear physical picture of adding
memory to a Maxwell Demon engine. If we imagine that the SMD has two bodies
$K_1$ and $K_2$ that store information, the single-particle Szilard engine can
operate for two cycles without having to go through an erasure process. Instead
of erasing the first body at the end of the first cycle, the SMD moves on to
operate on $K_2$---leaving $K_1$ in whatever final state the first cycle
determined. This avoids increasing the universe entropy while extracting work
from the Szilard engine heat bath. This violation of the Second Law is only
transient, though. To preform a third cycle, the SMD must erase $K_1$ or $K_2$
to store the next measurement. At the point immediately before erasure in each
cycle, the joint system must pay the entropic cost for $2$ fewer measurements
than it has made.

It is easy to see how this construction generalizes to larger physical memories
consisting of $N$ memory ``bits'' $K_1, K_2, \ldots, K_N$. With an $N$-bit
memory, the joint system of the Szilard engine and the SMD can continue to
extract work from the engine's reservoirs for $N$ cycles before having to
finally pay the cost of its first measurement. From that point forward, though,
every new measurement must be associated with an erasure of a previous one.
This restores the Second Law with respect to the erased measurement. Each
measurement is eventually paid for and, as the number of cycles grows large,
the transient leverage from having a large memory becomes less and less
noticeable.

\section{Conclusion}
\label{sec:Conclusion}

Since Szilard's day in 1929, the once-abstract conception of a molecular-scale
``neat fingered and very observant'' being \cite{Maxw88a} that interacts with
heat and information reservoirs has only become more tenable, as modern
computing emerged and micromanipulators were invented and then miniaturized
through nanofabrication techniques. Thus, understanding the workings of
information engines---microscopic machines interacting with such
reservoirs---is now highly relevant, especially compared to the days when
Maxwell first offered up the idea as a pedagogical absurdity. This is evinced
by, if nothing else, a constant and increasing stream of recent efforts that
take Szilard's original single-molecule engine as a jumping off point to
investigate how measurement, information, thermodynamics, and energy interact
with one another in support functional behaviors \cite{still2020thermodynamic,
bengtsson2018quantum, mohammady2017quantum, vaikuntanathan2011modeling,
kish2012energy, zurek2018eliminating}.

Szilard's early models grounded Maxwell's demon in physical embeddings. Since
their introduction they provided the bedrock for much debate and occasional
insights over the past century, largely through his first, brilliantly-simple
single-particle engine. Here, we found that his second (multiparticle) engine,
though more obtuse in construction, captures all of the same interesting
consequences suggested by the first. Additionally, it maps exactly on to the
first engine's operation by setting the demon memory state to another
positional coordinate as in Ref. \cite{Boyd14b}. In several important ways,
though, his second engine is more physical and plausible. And so, the
multiparticle-membrane engine is more robust to criticism arising from concerns
about applying classical statistics to the behavior of the first engine's
single particle. Thus, the second engine's relationship to its single-particle
sibling supports the physicality of the limits on information costs as
developed in Refs. \cite{Benn82,Land61a,Parrondo2015,Boyd16d,Zasl95a}.

Here, we re-emphasized the connection between Maxwell's demon and deterministic
chaos. This has been discussed in several settings
\cite{Boyd14b,Zasl95a,Seifert2012,Parrondo2001}. Reference \cite{Boyd14b}, for
example, noted that the degree of chaos---the Kolmogorov-Sinai entropy of the
engine's equivalent chaotic dynamical system---is the rate at which the engine
(transiently) extracts disorganized thermal energy transforming it into work.
In this way, the dynamical-systems connection provides a powerful approach to
precisely accounting for the simultaneous flow of energy and information.
Indeed, the connection is fated. A device that exhibits demon-like behavior is
a machine that takes microscopic thermal fluctuations and amplifies them to
macroscopic effect. A deterministic chaotic system is one in which microscopic
variations in initial conditions yield macroscopic variation in its trajectory.
With this in mind and reflecting on the century-plus history of clever
constructions of Maxwellian demons, the current constructions appear especially
suited to deepening our understanding of the relationship between energy and
information.

\section{Looking Ahead}

Beyond the conceptual insights that arise from Szilard's various engines, we
can even be somewhat literal-minded. Szilard's first engine has been the
inspiration for a diverse set of models and experimental realizations
\cite{admon2018experimental, koski2014experimental, koski2014experimental2,
koski2016maxwell, kish2012electrical}. Recent developments in nanofabrication
suggest attempting to realize Szilard's multi-particle engine, as well. For
example, the graphene membrane fabrication techniques discussed in Ref.
\cite{choi2018membrane} can provide macroscale membranes with tunable pore
size, pore density, and mechanical strength that are well suited to molecular
gas separation. With the right gas ensemble, it is possible these are good
candidates for the semi-permeable membranes required by Szilard's second
engine. When coupled with modern nanomechanical device design, a tantalizing
engineering challenge to implement Szilard's second engine presents itself.

Szilard's engines are simple enough to be readily analyzed, as we showed, with
all hitherto relevant thermodynamic and information calculations analytically
solvable. This thorough-going look at Szilard's original constructions gave
reassuring results---results consistent with the fundamentals of both
information theory and thermodynamics. Curiously, they are also consistent with
very recent developments in nonequilbrium thermodynamics and fluctuation
theory.

In particular, our investigation raised new questions. For one, Szilard's
inequality $e^{-S_1 /k} + e^{-S_2 /k} \leq 1 $ is more akin to very modern
fluctuation theorems \cite{Sagawa2012,Seifert2012,Jarzynski2011,Crooks2016}
than to the fluctuation theories of his contemporaries. Is this another realm
in which Szilard was prescient? This would not be surprising given that Szilard
anticipated Shannon's information theory by two decades, Wiener's cybernetics,
and the rise and fall of Landauer's Principle by half a century. We speculate
that Szilard's constructions can again provide a simple platform---one giving a
new view of detailed fluctuation theorems in action.

\section*{Acknowledgments}
\label{sec:acknowledgments}

We thank Alec Boyd for helpful discussions and the Telluride Science Research
Center for their hospitality during visits. This material is based upon work
supported by, or in part by, FQXi Grants FQXi-RFP-1609 and FQXi-RFP-IPW-1902,
the John Templeton Foundation grant 52095, and U.S. Army Research Laboratory
and the U. S. Army Research Office under contracts W911NF-13-1-0390 and
W911NF-18-1-0028.

\bibliography{Szilard2,chaos,ref}

\begin{thebibliography}{10}

\bibitem{Maxw88a}
J.~C. Maxwell.
\newblock {\em Theory of Heat}.
\newblock Longmans, Green and Co., London, United Kingdom, ninth edition, 1888.

\bibitem{Thomson1874}
W.~Thomson.
\newblock Kinetic theory of the dissipation of energy.
\newblock {\em Nature}, 9:441 EP --, 04 1874.

\bibitem{Leff02a}
H.~Leff and A.~Rex.
\newblock {\em Maxwell's Demon 2: {Entropy}, Classical and Quantum Information,
  Computing}.
\newblock Taylor and Francis, New York, 2002.

\bibitem{Thomson1879}
W.~Thomson.
\newblock The sorting demon of {Maxwell}.
\newblock In {\em Roy. Soc. Proc.}, volume~9, pages 113--114, 1879.

\bibitem{Szil29a}
L.~Szilard.
\newblock On the decrease of entropy in a thermodynamic system by the
  intervention of intelligent beings.
\newblock {\em Z. Phys.}, 53:840--856, 1929.

\bibitem{Lano13a}
W.~Lanouette and B.~Szilard.
\newblock {\em Genius in the Shadows: {A} Biography of Leo Szilard, {The} Man
  Behind The Bomb}.
\newblock Skyhorse Publishing, New York, New York, 2013.

\bibitem{Landauer1961}
R.~Landauer.
\newblock Irreversibility and heat generation in the computing process.
\newblock {\em IBM J. Res. Dev.}, 5(3):183--191, July 1961.

\bibitem{Benn82}
C.~H. Bennett.
\newblock Thermodynamics of computation---{A} review.
\newblock {\em Intl. J. Theo. Phys.}, 21:905, 1982.

\bibitem{Land61a}
R.~Landauer.
\newblock Irreversibility and heat generation in the computing process.
\newblock {\em IBM J. Res. Develop.}, 5(3):183--191, 1961.

\bibitem{Shiz95a}
K.~Shizume.
\newblock Heat generation required by information erasure.
\newblock {\em Phys. Rev. E}, 52(4):3495--3499, 1995.

\bibitem{Fahn96a}
F.~N. Fahn.
\newblock {Maxwell's} demon and the entropy cost of information.
\newblock {\em Found. Physics}, 26:71--93, 1996.

\bibitem{Bark06a}
M.~M. Barkeshli.
\newblock Dissipationless information erasure and {Landauer's} principle.
\newblock {\em arXiv:0504323}.

\bibitem{Saga12a}
T.~Sagawa.
\newblock Thermodynamics of information processing in small systems.
\newblock {\em Prog. Theo. Phys.}, 127(1):1--56, 2012.

\bibitem{Boyd14b}
A.~B. Boyd and J.~P. Crutchfield.
\newblock Maxwell demon dynamics: {Deterministic} chaos, the {Szilard} map, and
  the intelligence of thermodynamic systems.
\newblock {\em Phys. Rev. Lett.}, 116:190601, 2016.

\bibitem{Path96a}
R.~K. Pathria and P.~D. Beale.
\newblock {\em Statistical Mechanics}.
\newblock Butterwork-Heinemann, Oxford, United Kingdom, second edition, 1996.

\bibitem{Shan48a}
C.~E. Shannon.
\newblock A mathematical theory of communication.
\newblock {\em Bell Sys. Tech. J.}, 27:379--423, 623--656, 1948.

\bibitem{Cover2012}
T.~M. Cover and J.~A. Thomas.
\newblock {\em Elements of Information Theory}.
\newblock John Wiley \& Sons, 2012.

\bibitem{Seifert2012}
U.~Seifert.
\newblock Stochastic thermodynamics, fluctuation theorems and molecular
  machines.
\newblock {\em Reports Prog. Phys.}, 75(12):126001, 2012.

\bibitem{Parrondo2015}
J.~M.~R. Parrondo, J.~M. Horowitz, and T.~Sagawa.
\newblock Thermodynamics of information.
\newblock {\em Nature Physics}, 11(2):131, 2015.

\bibitem{Laso85a}
A.~Lasota and M.~C. Mackey.
\newblock {\em Probabilistic Properties of Deterministic Systems}.
\newblock Cambridge University press, Cambridge, United Kingdom, 1985.

\bibitem{Crut12a}
J.~P. Crutchfield.
\newblock Between order and chaos.
\newblock {\em Nature Physics}, 8(January):17--24, 2012.

\bibitem{Stro94a}
S.~H. Strogatz.
\newblock {\em Nonlinear Dynamics and Chaos: with applications to physics,
  biology, chemistry, and engineering}.
\newblock Addison-Wesley, Reading, Massachusetts, 1994.

\bibitem{Barn13a}
N.~Barnett and J.~P. Crutchfield.
\newblock Computational mechanics of input-output processes: {Structured}
  transformations and the $\epsilon$-transducer.
\newblock {\em J. Stat. Phys.}, 161(2):404--451, 2015.

\bibitem{Crut08b}
C.~J. Ellison, J.~R. Mahoney, and J.~P. Crutchfield.
\newblock Prediction, retrodiction, and the amount of information stored in the
  present.
\newblock {\em J. Stat. Phys.}, 136(6):1005--1034, 2009.

\bibitem{kolchinsky2017}
A.~Kolchinsky and D.~H Wolpert.
\newblock Dependence of dissipation on the initial distribution over states.
\newblock {\em J. Stat. Mech.: Th. Expt.}, 2017(8):083202, 2017.

\bibitem{Ashb57a}
W.~Ross Ashby.
\newblock {\em An Introduction to Cybernetics}.
\newblock John Wiley and Sons, New York, second edition, 1960.

\bibitem{Boyd16d}
A.~B. Boyd, D.~Mandal, and J.~P. Crutchfield.
\newblock Leveraging environmental correlations: The thermodynamics of
  requisite variety.
\newblock {\em J. Stat. Phys.}, 167(6):1555--1585, 2016.

\bibitem{still2020thermodynamic}
S.~Still.
\newblock Thermodynamic cost and benefit of memory.
\newblock {\em Phys. Rev. Let.}, 124(5):050601, 2020.

\bibitem{bengtsson2018quantum}
J.~Bengtsson, M.~N. Tengstrand, A.~Wacker, P.~Samuelsson, M.~Ueda, H.~Linke,
  and S.~M. Reimann.
\newblock Quantum {Szilard} engine with attractively interacting bosons.
\newblock {\em Phys. Rev. Let.}, 120(10):100601, 2018.

\bibitem{mohammady2017quantum}
M.~H. Mohammady and J.~Anders.
\newblock A quantum {Szilard} engine without heat from a thermal reservoir.
\newblock {\em New J. Physics}, 19(11):113026, 2017.

\bibitem{vaikuntanathan2011modeling}
S.~Vaikuntanathan and C.~Christopher.
\newblock Modeling {Maxwell's} demon with a microcanonical {Szilard} engine.
\newblock {\em Phys. Rev. E}, 83(6):061120, 2011.

\bibitem{kish2012energy}
L.~B. Kish and C.~G. Granqvist.
\newblock Energy requirement of control: Comments on {Szilard's} engine and
  {Maxwell's} demon.
\newblock {\em Europhys. Lett.}, 98(6):68001, 2012.

\bibitem{zurek2018eliminating}
W.~H. Zurek.
\newblock Eliminating ensembles from equilibrium statistical physics:
  {Maxwell's} demon, {Szilard's} engine, and thermodynamics via entanglement.
\newblock {\em Phys. Rep.}, 755:1--21, 2018.

\bibitem{Zasl95a}
G.~M. Zaslavsky.
\newblock From {Hamiltonian} chaos to {Maxwell's} demon.
\newblock {\em Chaos}, 5:653--661, 1995.

\bibitem{Parrondo2001}
J.~M.~R. Parrondo.
\newblock The {Szilard} engine revisited: Entropy, macroscopic randomness, and
  symmetry breaking phase transitions.
\newblock {\em Chaos: Interdisc. J. Nonlin. Sci.}, 11(3):725--733, 2001.

\bibitem{admon2018experimental}
T.~Admon, S.~Rahav, and Y.~Roichman.
\newblock Experimental realization of an information machine with tunable
  temporal correlations.
\newblock {\em Phys. Rev. Let.}, 121(18):180601, 2018.

\bibitem{koski2014experimental}
J.~V. Koski, V.~F. Maisi, J.~P. Pekola, and D.~V. Averin.
\newblock Experimental realization of a {Szilard} engine with a single
  electron.
\newblock {\em Proc. Natl. Acad. Sci. USA}, 111(38):13786--13789, 2014.

\bibitem{koski2014experimental2}
K.~V. Koski, V.~F. Maisi, T.~Sagawa, and J.~P. Pekola.
\newblock Experimental observation of the role of mutual information in the
  nonequilibrium dynamics of a {Maxwell} demon.
\newblock {\em Phys. Rev. Let.}, 113(3):030601, 2014.

\bibitem{koski2016maxwell}
J.~V. Koski and J.P. Pekola.
\newblock {Maxwell's} demons realized in electronic circuits.
\newblock {\em Compt. Rend. Phys.}, 17(10):1130--1138, 2016.

\bibitem{kish2012electrical}
L.~B. Kish and C.-G. Granqvist.
\newblock Electrical {Maxwell} demon and {Szilard} engine utilizing {Johnson}
  noise, measurement, logic and control.
\newblock {\em PloS One}, 7(10), 2012.

\bibitem{choi2018membrane}
K.~Choi, A.~Droudian, R.~W. Wyss, K.-P. Schlichting, and H.~G. Park.
\newblock Multifunctional wafer-scale graphene membranes for fast
  ultrafiltration and high permeation gas separation.
\newblock {\em Sci. Adv.}, 4(11):eaau0476, 2018.

\bibitem{Sagawa2012}
T.~Sagawa and M.~Ueda.
\newblock Fluctuation theorem with information exchange: Role of correlations
  in stochastic thermodynamics.
\newblock {\em Phys. Rev. Lett.}, 109(18):180602, 2012.

\bibitem{Jarzynski2011}
C.~Jarzynski.
\newblock Equalities and inequalities: Irreversibility and the second law of
  thermodynamics at the nanoscale.
\newblock {\em Annu. Rev. Condens. Matter Phys.}, 2(1):329--351, 2011.

\bibitem{Crooks2016}
G.~E. Crooks and S.~E. Still.
\newblock Marginal and conditional second laws of thermodynamics.
\newblock {\em arXiv:1611.04628}.

\end{thebibliography}

\appendix 

\section{Entropy Change}
\label{app:EntropyChange}

Our goal is to determine $\Delta S$ in Szilard's second engine. The
Sakur-Tetrode equation, the starting point, is:
\begin{align*}
S = N\kB \ln \frac{V}{N} + \frac{3}{2} N \kB \ln \frac{4\pi m U}{3h^2N} +\frac{5}{2} N \kB
  ~.
\end{align*}
Terms that remain constant throughout an engine cycle can be neglected for our
purposes. Cursory inspection reveals that $\Delta S$ will be determined by, at
most:
\begin{align*}
N \kB \ln \frac{V}{N} + \frac{3}{2} N \kB \ln \frac{U}{N} 
  ~.
\end{align*}
In our case, the energy density term also drops out, since both the initial and
final macrostates reach the equilibrium distribution $N_\TSquare = \delta N$
and $N_\SubTCircle = (1-\delta) N$. And so:
\begin{align*}
\frac{U}{N} = \delta \epsilon_A + (1-\delta)\epsilon_B + \KE
  ~,
\end{align*}
for any number of particles, where $\KE$ is the average kinetic energy.
Finally, since each container has the same volume, the volume term does not
contribute. Calculating the resulting entropy change is straightforward,
but requires attention. After some algebra, we have:
\begin{align*}
\frac{\Delta S}{\kB}
  & = - \delta \ln N_\TSquare - (1-\delta) \ln N_\SubTCircle + \ln N \\
  & = -  \left( \delta \ln  \delta + 1-\delta \ln (1-\delta)\right) \\
  & = S(\delta) 
  ~.
\end{align*}

\section{Free Energies}
\label{app:FreeEnergies}

A similar need arises for obtaining the free energy in Szilard's second
engine. We starting observing that:
\begin{align*}
 \Wqs = \Delta \Ures + \Delta \Usys 
  ~.
\end{align*}
If reservoir volume remains constant, we note that $\Qres = \Delta \Ures$.
Then, using the expression above, we find $\Delta \Sres = \Qres / T$ to be
given by:
\begin{align*}
T \Delta \Sres = \Wqs - \Delta \Usys
  ~.
\end{align*}
Thus, the total entropy change, including both the system and the reservoir
must then be:
\begin{align*}
T\Delta \Sres + T\Delta \Ssys = \Wqs + T \Delta \Ssys - \Delta \Usys
  ~.
\end{align*}
The  difference of $T \Delta \Ssys - \Delta \Usys$ is nothing more than the
change in the system's free energy $-\Delta \Fsys$. Following Szilard's
statement, we choose a reversible process ($\Delta \Sres + \Delta \Ssys =0$).
This allows us to find the work to drive the quasistatic step by:
\begin{align*}
\Wqs = F(\rho_0) - F(\rho)
  ~.
\end{align*}
Adding $\Wqs$ and $W_{\Delta H}$ yields the the total driving work for both
steps of the coming to equilibrium process:
\begin{align*}
W_{drive} & = \langle H_\rho \rangle _\rho - \langle H_0 \rangle_\rho
  + \left( \langle H_0 \rangle_{\rho_0}  - TS(\rho_0) \right) \\
  & \qquad - \left(\langle H_\rho \rangle_\rho - TS(\rho)\right) \\
& = \langle H_0 \rangle_{\rho_0} - \langle H_0\rangle_\rho
  + T S(\rho) - T S(\rho_0)  
  ~.
\end{align*}

\iffalse
 We can also verify that its operation obeys the Second Law by
adding up the entropy production in the system and in the reservoir. 
\begin{align*}
\Delta \Stot & = \Delta \Ssys + \Delta \Sres \\
  & = \Delta \Ssys - \frac{Q_{drive}}{T} \\
  & = S(\rho_0) - S(\rho) + \frac{W_{drive}}{T} \\
  & = S(\rho_0) - S(\rho_0) \\
  & = 0 
  ~.
\end{align*}
Here, $Q_{drive} = -W_{drive}$ since there is no change in internal energy. We
see that, indeed, the process is reversible.
\fi

\section{An Erasure Alternative}
Consider a different choice for the final erasure step in the Szilard Map.

We could simply remove the partition and allow the gas to spontaneously
re-equilibrate. This gives the same relationship in terms of entropy, but we do
not get the advantage of a clear way to extract work that can be harnessed by
the entropy increase.

Specifically, the change in entropy when mixing two identical gasses at
different densities depends only on that part of the entropy given by $ \ln
\frac{V}{N}$. Initially, there are two separate gasses with the relevant entropy
components:
\begin{align*}
S_A + S_B = N \kB \delta \ln \frac{\ell\gamma}{\delta N} + N \kB (1-\delta) \ln
\frac{\ell(1-\gamma)}{(1-\delta)N}
  ~.
\end{align*}
In the final state, we have a single gas:
\begin{align*}
S_F = N \kB \ln \frac{\ell}{N}
  ~.
\end{align*}
The difference gives the entropy change:
\begin{align*}
\frac{\Delta S}{N \kB} &= \ln \frac{1}{N}
  - \delta \ln \frac{\gamma}{\delta N}
  - (1-\delta) \ln \frac{1-\gamma}{(1-\delta)N} \\
  & = - \left( (1-\delta) \ln \frac{1-\gamma}{1-\delta} + \delta \ln \frac{\gamma}{\delta}\right)
  ~.
\end{align*}

\section{Szilard Engine Maps}
\label{app:DeterministicMaps}

The three discrete-time maps of the unit-square state-space that correspond
to \emph{measurement}, \emph{control}, and \emph{erasure} are, respectively:
\begin{align*}
\tau_{\text{M}} (\PTyp,\PMem) = \begin{cases} 
      \left(\PTyp, \PMem \gamma \right) 			& \PTyp<\delta  \\
      \left(\PTyp, \PMem(1-\gamma)+\gamma\right) 	& \PTyp>\delta 
   \end{cases}
  ~,
\end{align*}
\begin{align*}
\tau_{\text{C}} (\PTyp,\PMem) = \begin{cases} 
      \left(\frac{\PTyp}{\delta}, \PMem \right) 			& \PMem<\gamma  \\
      \left(\frac{\PTyp-\delta}{1-\delta}, \PMem \right) 	& \PMem>\gamma
   \end{cases}
  ~,
\end{align*}
and:
\begin{align*}
\tau_{\text{E}} (\PTyp,\PMem) = \begin{cases} 
      \left(\PTyp, \frac{\PMem}{\gamma} \delta \right) 		& \PMem<\gamma  \\
      \left(\PTyp, \frac{(y-\gamma)(1-\delta)}{1-\gamma} + \delta \right) 	& \PMem>\gamma
   \end{cases}
  ~.
\end{align*}

Taken together, and specializing to the case where $\delta=\gamma$, we have the composite map:
\begin{align*}
\tau_{\text{Szilard}} (\PTyp,\PMem) = \begin{cases} 
      \left(\frac{\PTyp}{\delta}, \PMem \delta\right) & \PTyp<\delta  \\
      \left(\frac{\PTyp-\delta}{1-\delta}, \delta+\PMem(1-\delta)\right) & \PTyp>\delta 
   \end{cases}
  ~.
\end{align*}

\end{document}